\documentclass[useAMS,usenatbib]{mn2e}
\usepackage{graphicx}
\usepackage{latexsym}

\title[Fragmentation in Molecular Clouds]{Fragmentation in Molecular Clouds and its connection to the IMF}

\author[Smith, Clark \& Bonnell]{Rowan J. Smith$^{1}$ \thanks{Email: rjs22@st-andrews.ac.uk}, Paul C. Clark$^{2}$ and Ian A. Bonnell$^{1}$ \\
$^1$ SUPA, School of Physics \& Astronomy, University of St Andrews, North Haugh, St Andrews, Fife, KY16 9SS, UK \\
$^2$ Zentrum f\"ur Astronomie der Universit\"at Heidelberg, Institut f\"ur Theoretische Astrophysik,Albert-Ueberle-Str. 2, Heidelberg, Germany  
}

\begin{document}

\pagerange{\pageref{firstpage}--\pageref{lastpage}} \pubyear{2008}

\maketitle

\label{firstpage}

\def\solmas{{M$_\odot$}}
\def\solm{{M_\odot}}
\def\mnras{MNRAS}
\def\apj{ApJ}
\def\aap{A\&A}
\def\apjl{ApJL}
\def\apjs{ApJS}
\def\bain{BAIN}
\def\pasp{PASP}
\def\araa{ARA\&A}
\def\ga{\sim}

\begin{abstract}
We present an analysis of star-forming gas cores in an SPH simulation of a Giant Molecular Cloud. We identify cores using their deep potential wells. This yields a smoother distribution with clearer boundaries than density. Additionally, this gives an indication of future collapse, as bound potential cores (p-cores) represent the earliest stages of fragmentation in molecular clouds. We find that the mass function of the p-cores resembles the stellar IMF and the observed clump mass function, although p-core masses ($\sim0.7$ M$_{\odot}$) are smaller than typical density clumps. The bound p-cores are generally subsonic, have internal substructure, and are only quasi-spherical. We see no evidence of massive bound cores supported by turbulence. We trace the evolution of the p-cores forward in time, and investigate the connection between the original p-core mass and the stellar mass that formed from it. We find that there is a poor correlation, with considerable scatter suggesting accretion onto the core is dependent on more factors than just the initial core mass. During the accretion process the p-cores accrete from beyond the region first bound, highlighting the importance of the core environment to its subsequent evolution. 
\end{abstract}


\section{Introduction}
\label{sec:introduction}
The origin of the stellar initial mass function (IMF) has been a major question in star formation since it was first measured by \citet{Salpeter55}. The similarities between the IMF and the mass function of clumpy structure in molecular clouds (MC's), has led many to propose a link between the two (e.g. \citealp{Motte98}). In this scenario, stars are formed directly from the dense cores of gas observed in molecular clouds (e.g. \citealp{Johnstone00,Testi98,Johnstone06,Nutter07,Enoch08}).  A direct link between the core mass and the resulting stellar systems mass is often assumed. For example \citet{Alves07} propose an efficiency of one third between core mass and stellar system mass. 

This fits into the the long-standing idea that fragmentation of molecular cloud structure forms the IMF. This could be simply due to gravity \citep{Larson73,Elmegreen83}, causing successive fragmentation of a larger body as it collapses. Alternatively, it has been shown that the supersonic turbulence observed in MC's produces a hierarchical density structure, the dense peaks of which have a mass distribution similar to the IMF \citep{Henriksen86,Larson92,Elmegreen97,Klessen01,Padoan02,Hennebelle08} . The Jeans mass \citep{Jeans1902} at the point of fragmentation has been shown to be only weakly dependent on temperature, density, metalicity and radiation field in the environments in which stars form \citep{Larson05,Elmegreen08}, which leads to the characteristic mass of a Jeans unstable fragment being similar in all molecular clouds.

However, it is unclear how to get to the IMF from the clump mass function. Many assume a direct $1-1$ link between the masses \citep{Motte98,Padoan02,Alves07}, while others include the effects of multiplicity \citep{Goodwin08}. However, there are many complicating factors in this story of collapse, such as feedback from winds and outflows \citep{Shu88,Silk95,Myers08}, supporting magnetic fields \citep{Heitsch01,Tilley07} and competitive accretion \citep{Zinnecker82,Bonnell06}. All of these processes are involved in the collapse of a fragment to a star and all could vary locally. In fact, \citet{Swift08} have shown that when a core mass function is evolved into a stellar IMF, a Salpeter like distribution was found regardless of whether the core-to-star efficiency was constant, variable or included multiplicity. Moreover, under the competitive accretion theory of star formation there is no need for a direct correlation between core masses and stellar masses at all, as the cores can be thought of as `seeds' from which accretion will build up the future IMF \citep{Clark05}.


Resolving the issue of core evolution is further complicated as their lifetimes are not well known. In the classical \citet{Shu77} picture a quasi-static core supported by its magnetic field will slowly collapse to form a star. However, in the more dynamical view, usually proposed in gravoturbulent fragmentation, cores can collapse quickly when they become unstable, making the process hard to observe. Further, the variation of core free-fall time with density means that the clump mass function observed at a snap shot in time might actually evolve into a steeper IMF \citep{Clark07}.

Massive star formation poses another question for the evolution of cores. Any core which is large enough to form a massive star directly is likely to fragment without some additional heating mechanism. \citet{McKee03} theorize that massive cores could be supported by their internal turbulent energy, although simulations have shown that some fragmentation is unavoidable \citep{Krumholz07b,Dobbs05}.  Alternatively competitive accretion predicts that there will be no massive cores that do not fragment into smaller structures. It would be one of these smaller cores which would preferentially accrete from its environment to become a massive star \citep{Bonnell04}. In this case, the link between core mass and stellar mass is destroyed by accretion.

Observationally it is hard to be sure which structures will gravitationally decouple from their environment and are hence `pre-stellar' in nature (i.e. will form stars in their future). Observations of molecular tracers such as CO produce a core mass function resembling the IMF ( e.g. \citealp{Ikeda07}). However, these tracers are insensitive to the densest gas. As \citet{Lada_E92} has shown, star formation generally takes place above a density of $n>10^{4}$ cm$^{-3}$. Observations of CO cores are typically more massive and on larger scales ( e.g. \citealp{Tachihara02}) than mm-continuum cores which trace denser gas, ( e.g. \citealp{Motte98,Johnstone00}). Moreover, it is often difficult to define a `core' from the data, without invoking a somewhat arbitrary boundary (e.g. \citealp{Padoan06,Pineda08,Schnee08}). In particular, \citet*{Smith08} demonstrated that core properties are extremely sensitive to the core boundaries, which in turn depend on the resolution, density range and dimensionality of the dataset. High resolution observations of dense gas in nearby molecular clouds are the most likely to find pre-stellar objects, for example those identified from a synthesis and re-analysis of the literature in Ophiuchus by \citet{Simpson08}. However, these observations still suffer from a lack of completeness, and of course all observations are necessarily at a snap-shot in time and identify cores in a variety of evolutionary states with no guarantee that they will form stars.

In this work we seek to determine the relation between the properties of the cores formed through the fragmentation of molecular clouds and the `stars' which form from them. In our SPH simulation of a molecular cloud we identify bound pre-stellar cores with well defined boundaries in a complete data-set without time effects. Our cores are gravitationally bound, and as they cannot be temporarily confined by pressure they will collapse (c.f. \citealt{Dib07}). We trace the core evolution from when first bound through their early evolution towards stars. Section \ref{sec:simulation} outlines the simulation initial conditions, and Section \ref{sec:clumpfind} discusses how we define our cores, here called p-cores, using their gravitational potential to ensure they are bound with respect to their environment. In Section \ref{sec:physical} we calculate the p-core properties, compare them with observations and find the core mass function. In Section \ref{sec:masses} we link the p-core masses when they are first bound with the mass which will be accreted by the stars (sink particles). Finally in Section \ref{sec:discussion} we discuss our results.


\section{The Simulation}
\label{sec:simulation}

We use a three dimensional smooth particle hydrodynamics (SPH) code \citep{Monaghan92}, to simulate star formation in a collapsing gas cloud. The SPH code has variable smoothing lengths, artificial viscosity \citep{Gingold83} and pressure gradients. Self-gravity is calculated via a binary tree \citep{Benz90} and sink particles \citep{Bate95} are used to model the star formation and prevent the simulation exceeding its mass resolution. 

The initial conditions consisted of a cylinder containing $10^{4}$ M$_{\odot}$ concentrated at one end so the top was over-bound and the bottom under-bound. A turbulent velocity grid, consistent with a Larson velocity dispersion of $\sigma \propto r^{0.5}$, was generated according to \citet{Dubinski95} and \citet{Myers99} and interpolated onto the SPH particles. The magnitude of the turbulent velocities is chosen such that globally the cloud is initially supported by turbulence, which equates to a r.m.s. velocity of $4.7$ kms$^{-1}$. The turbulence is not driven but additional kinetic energy is released by the gravitational collapse of the MC, when the simulation was terminated the r.m.s. velocity had decayed to $3.65$ kms$^{-1}$.

Table \ref{simprops} shows the properties of the simulation.The cloud is modelled with $5.5 \times 10^{6}$ SPH particles, which gives a mass resolution of $0.18$ M$_{\odot}$ \citep{Bate97}. The simulation was run on the SUPA Altix computer at The University of St Andrews.

\begin{table}
	\centering
		\caption{Given below are the initial conditions of the simulation analysed in this paper. The mass resolution is the minimum mass gravitational forces can be resolved for and is calculated via $M_{res} \sim 100 M_{total}/N_{part}$.}
		\begin{tabular}{l c }
   	         \hline
	         \hline
	         Size & $3 \times 3 \times 10$ pc \\
	         Mass & $10^{4}$ M$_{\odot}$\\
	         Particles & $5.5\times10^{6}$\\
	         Mass resolution & $0.18$ M$_{\odot}$\\
	         Dynamical time & $4.74\times 10^{5} $ yrs\\
		\hline
		\end{tabular}
	\label{simprops}
\end{table}

A barotropic equation of state is used for basic heating and cooling to ensure that the Jeans mass at the point of fragmentation matches the characteristic stellar mass.
\begin{equation}
P=k\rho^{\gamma}
\end{equation}
where
\begin{equation}
\begin{tabular}{l l l }
$\gamma=0.75:$&$\rho\leq\rho_{1}$ & line cooling\\
$\gamma=1.0:$&$\rho_{1}\leq\rho\leq\rho_{2}$ & dust cooling \\
$\gamma=1.4:$&$\rho_{2}\leq\rho\leq\rho_{3}$ & optically thick to IR\\
$\gamma=1.0:$&$\rho\geq\rho_{3}$ & allow sink formation\\
\end{tabular}
\end{equation}
and $\rho_{1}=5.5\times10^{-19}$ gcm$^{-3}$,$\rho_{2}=5.5\times10^{-15}$ gcm$^{-3}$ and $\rho_{3}=2\times10^{-13}$ gcm$^{-3}$.

This equation of state mimics the effects of line cooling \citep{Larson05,Jappsen05} and then dust cooling when the dust is coupled to the gas \citep{Masunaga00}. When the gas becomes optically thick to IR radiation the gas will heat again. We invoke this heating at somewhat earlier stage in the collapse than typical ({$\rho_{2}=5.5\times10^{-15}$ gcm$^{-3}$}) to ensure the Jeans mass of a fragment is always resolved. Finally, we return to an isothermal equation of state at $\rho_{3}=2\times10^{-13}$ gcm$^{-3}$, to allow sink particles to form, as we require the fragments from which they form to be both bound and collapsing. Therefore in our simulation, we will not create any sink particles until at least a density of $\rho_{3}=2\times10^{-13}$ gcm$^{-3}$ has been reached. Sink particles will accrete bound material within a radius of $200$ AU, and have their mutual gravitational interactions smoothed to $40$ AU.

\begin{figure}
\begin{center}
\includegraphics[width=2.5in]{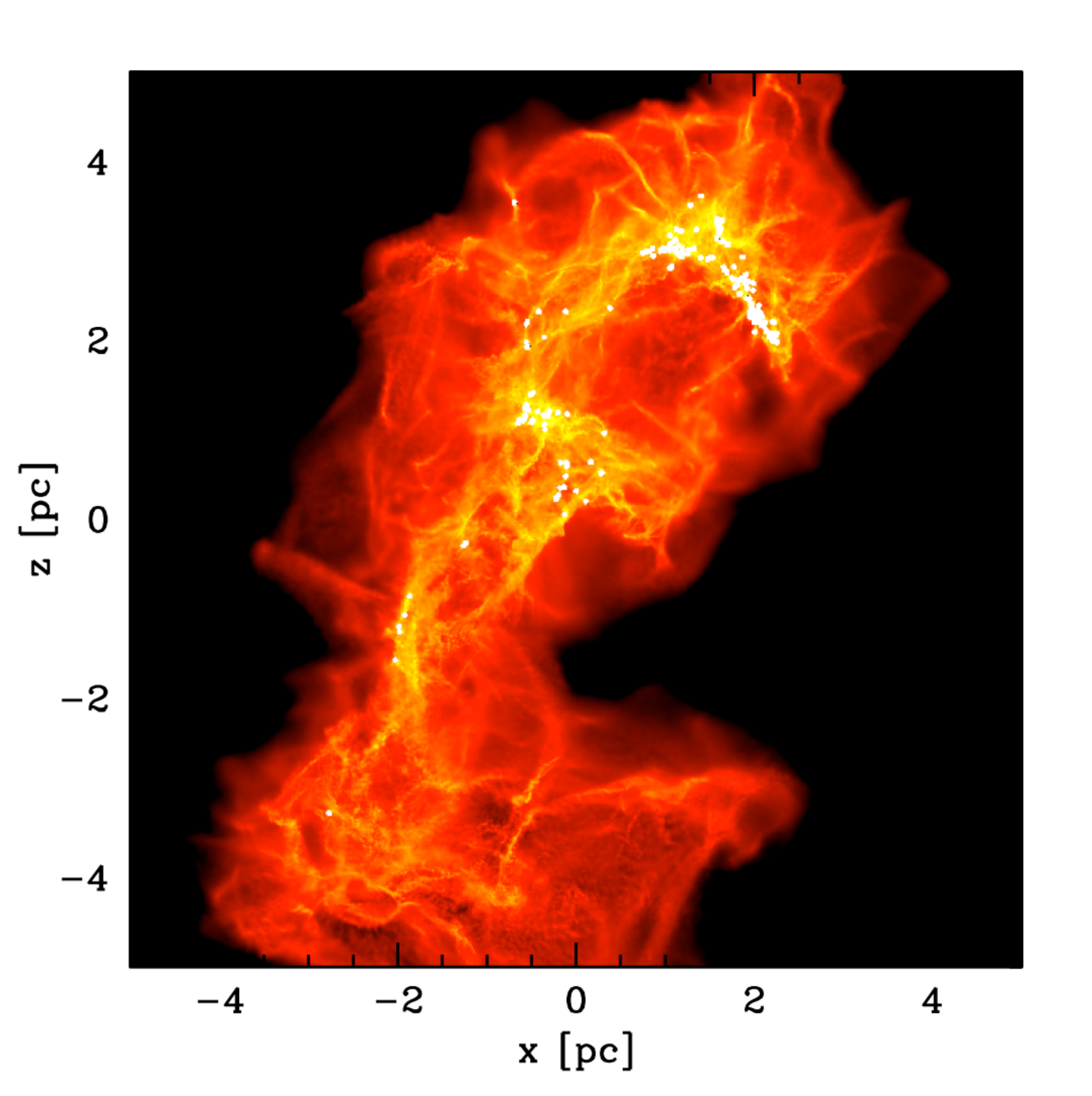}
\caption{The simulated Giant Molecular Cloud. The colours represent column densities in the range 0.001gcm$^{-2}$ (red) to 10 gcm$^{-2}$ (white). Sink particles are shown as white dots and the cloud is viewed along its  long axis.}
\label{standard}
\end{center}
\end{figure}

The simulation evolves with self gravity until just after a dynamical time it is as shown in Figure \ref{standard}. The simulation is terminated at $1.4$ t$_{dyn}$, at which point there are $949$ sink particles which have a combined mass of $1249$ M$_{\odot}$. This equates to a star formation efficiency of $12.5\%$.

\section{Clump Finding using Potential}
\label{sec:clumpfind}

As previously shown in \citet{Smith08}, clump mass functions always retain a universal profile, but the sizes, masses and positions of cores vary with the method of extraction. This behaviour is also seen in observations. \citet{Motte98} and \cite{Johnstone00} in their observations of $\rho$ Ophiuchus find similar clump mass functions but the positions of their clumps and their masses do not correspond with each other. As the density distribution of molecular clouds is hierarchical with structure on all scales it is unclear which scale is the most relevant for star formation. 

In this study, we take a different approach, since we are interested in establishing the connection between the mass of the fragment as it first becomes bound and the final mass of the sink particle that forms from it. Rather than breaking the cloud up into structures defined by density, we instead look at peaks in the gravitational potential. There are two advantages to such a method. First, the gravitational potential distribution in the cloud is considerably smoother than the density distribution, since density fluctuations that do not carry sufficient mass cannot significantly contribute to potential field. Second, the strength of the gravitational potential determines whether a clump will collapse and how mass will flow. In a density distribution it is unclear which scale of the structure is important, whereas with potential the scale at which structures are bound is a clear physical quantity. Naturally, there is a disadvantage to this process: it becomes difficult to compare our structures to those observed in molecular clouds. However we believe some properties may be comparable between the observed objects and those which are extracted from the simulation in this study. We discuss these in Section \ref{sec:physical}.

\begin{figure}
\begin{center}
\includegraphics[width=3in]{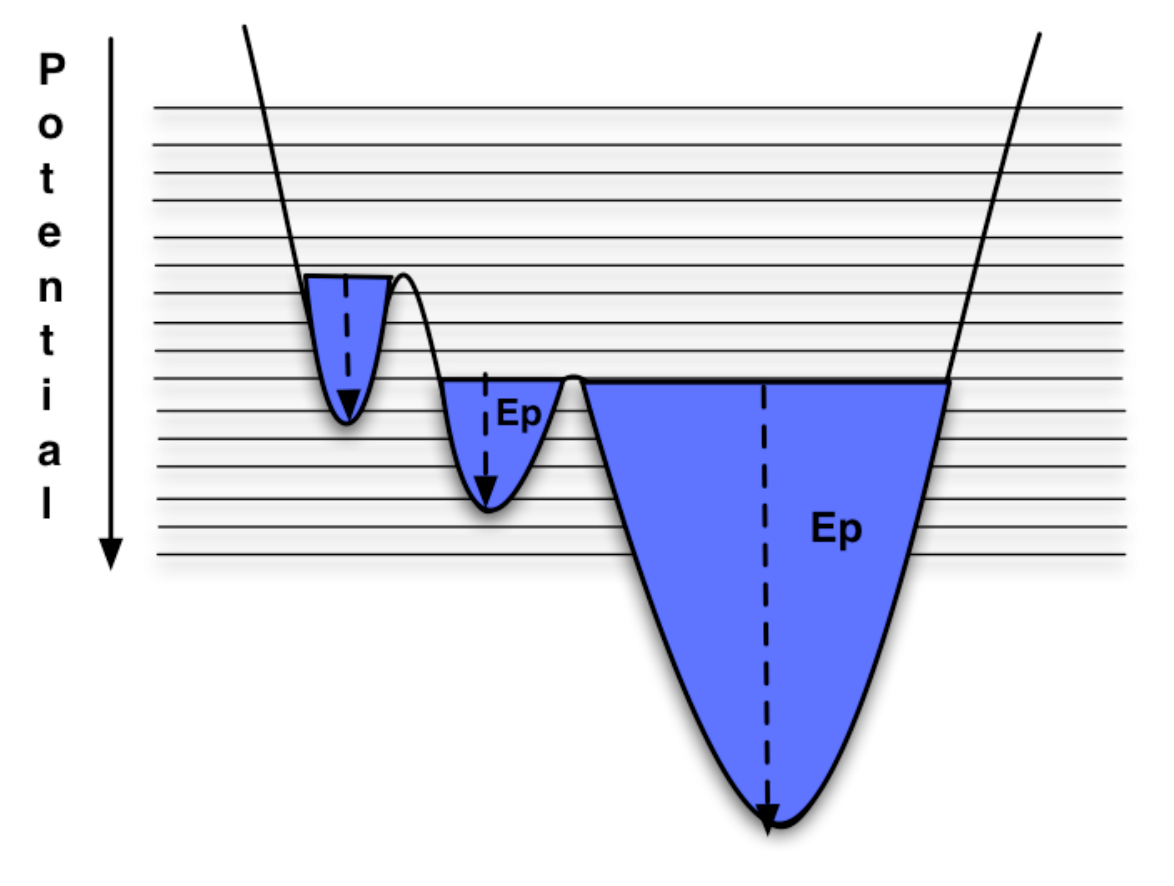}
\caption{A cartoon of the potential clump finding process in 1D. For the potential and contours shown, the blue regions contain material which would be assigned to p-cores. }
\label{cartoon}
\end{center}
\end{figure}

Potential wells are identified using an algorithm similar to CLUMPFIND \citep{Williams94}, which works directly on SPH particles \citep{Klessen00}. We have modified the algorithm to use gravitational potential to identify clumps rather than emission or density. In this scheme the SPH particle with the deepest gravitational potential forms the head of a clump, then the particle with the next deepest potential is either assigned to the same clump if it is a neighbour, or forms a new clump if not, and so on. Clumps are defined down to either a minimum positive potential, or the lowest contour which it shares with a neighbouring clump. Unlike the traditional CLUMPFIND algorithm, we use contour levels primarily to define the level at which potential clumps join, rather than to distinguish clumps from noise, and so our contours are numerous and finely spaced. This has the effect of subtracting the background potential. This is necessary as gravity is a long range force, and so is affected by both the mass inside and surrounding the p-core; hence we must remove the background to obtain the net effect on the mass within. P-cores, therefore, represent the local maximum above the surrounding background. Figure \ref{cartoon} shows a simple 1D cartoon of the potential clump finding algorithm where three potential clumps have been identified. Figure \ref{exampleclumps} shows an example of the potential clumps found from the central region of Figure \ref{standard}.  
 
It is now wise to define our future terminology. Usually (though not always) a `clump' is said to contain $50-500$ M$_{\odot}$ within $0.3-3$ pc, and a `core' is said to contain $0.5-5$ M$_{\odot}$ within $0.03-0.2$ pc \citep{Bergin07}. Following this convention, as the structures identified by the potential clump find are typically small, we shall call them `p-cores'. Note that there is no intrinsic requirement that a p-core is bound, as it may have sufficient internal energy (both thermal and kinetic) to prevent collapse. Once our p-cores form sinks, we only trace the total sink mass formed from them, because the spike in potential due to the sink particle distorts the boundaries; therefore all our p-cores are pre-stellar.

The major aim of this paper is to investigate the link between pre-stellar p-cores and the stars formed from them. In order to do this we use two data-sets, the first of which is the `composite' data-set. P-cores are found at $0.1$ dynamical time ($t_{dyn}=4.7\times10^{5}$ yrs) intervals between $0.6$ $t_{dyn}$ when star formation had just begun and $1.4$ $t_{dyn}$ when the simulation was terminated. Snapshots from each time are combined into one data-set, which removes any time dependent effects and increases our data-set. The cores in the composite data-set have an average dynamical time of $7.4\times10^{4}$ yrs meaning that a few long lived, non-transient cores will be included twice, but at different evolutionary states. In total there are $573$ p-cores. The composite data-set represents the view of cores in a molecular cloud at single points in time. It contains p-cores at different evolutionary states, with different levels of binding, many of which are transient.

The second data-set is the `bound' data-set, which contains the details of the p-cores at the point they first became bound. The p-cores are traced throughout the lifetime of the simulation, and if more than $80\%$ of the mass belonging to a p-core remains grouped together in the next simulation time-step, the p-core is said to survive. The binding is traced throughout the p-core lifetime, and at the point where $E_{rat}>1$ for the first time its properties are recorded. $E_{rat}$ is defined as:
\begin{equation}
\label{Erat}
E_{rat}=\frac{|E_{p}|}{E_{therm}+E_{k}}
\end{equation}
where $E_{k}$ is the kinetic energy calculated with respect to the center of velocity of the clump, $E_{therm}$ is the thermal energy of the clump and $E_{p}$ is the potential energy of the clump calculated using the relative depth of the potential well once the background has been subtracted. Hence the clumps identified here are bound with respect to the environment in which they are formed, not merely when considered in isolation. This means that tidal forces from surrounding cores are taken into account when determining whether the p-core is bound. As the bound p-cores are identified throughout the simulation this data-set is also time independent. Due to this data-set being a synthesis over time and the p-cores being identified using a quantity impossible to observe, these objects could not be found observationally. Despite this, the analysis is worthwhile as the bound data-set allows us to identify the fragmentation scale of the Molecular Cloud. The sink particles form when these bound cores collapse. In total there are $306$ bound cores in our simulation.


\section{Physical Properties of the Potential Cores}
\label{sec:physical}
Unfortunately, a potential core can only be identified when positions and velocities are known in three dimensions, which is impossible observationally. Therefore, we now calculate the properties of the p-cores to allow a comparison to observational data. The average properties are shown in Table \ref{cprops}. 

\subsection{P-core Shapes}

\begin{figure*}
\begin{center}
\begin{tabular}{c c}
\includegraphics[width=3in]{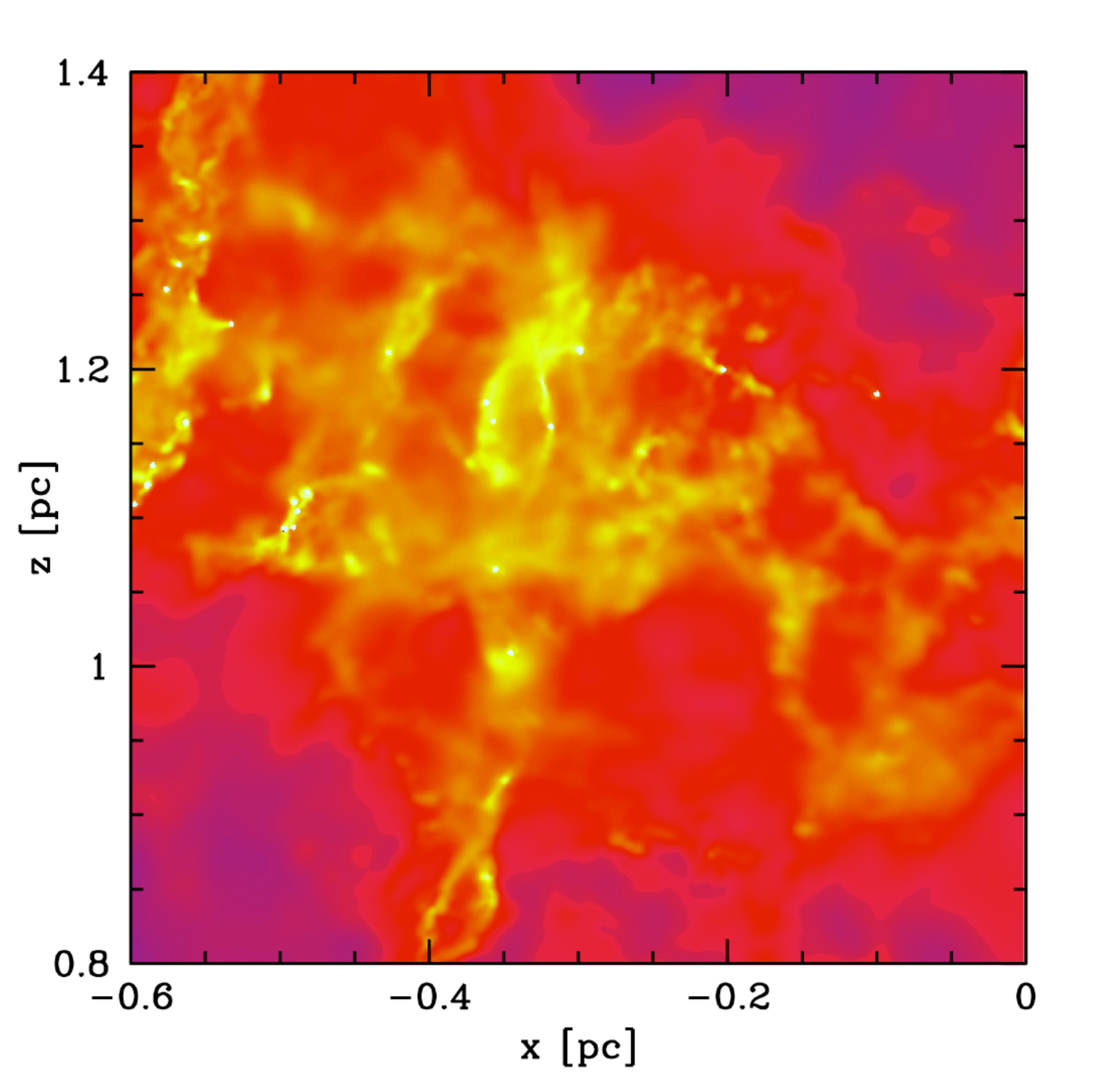} &
\includegraphics[width=3in]{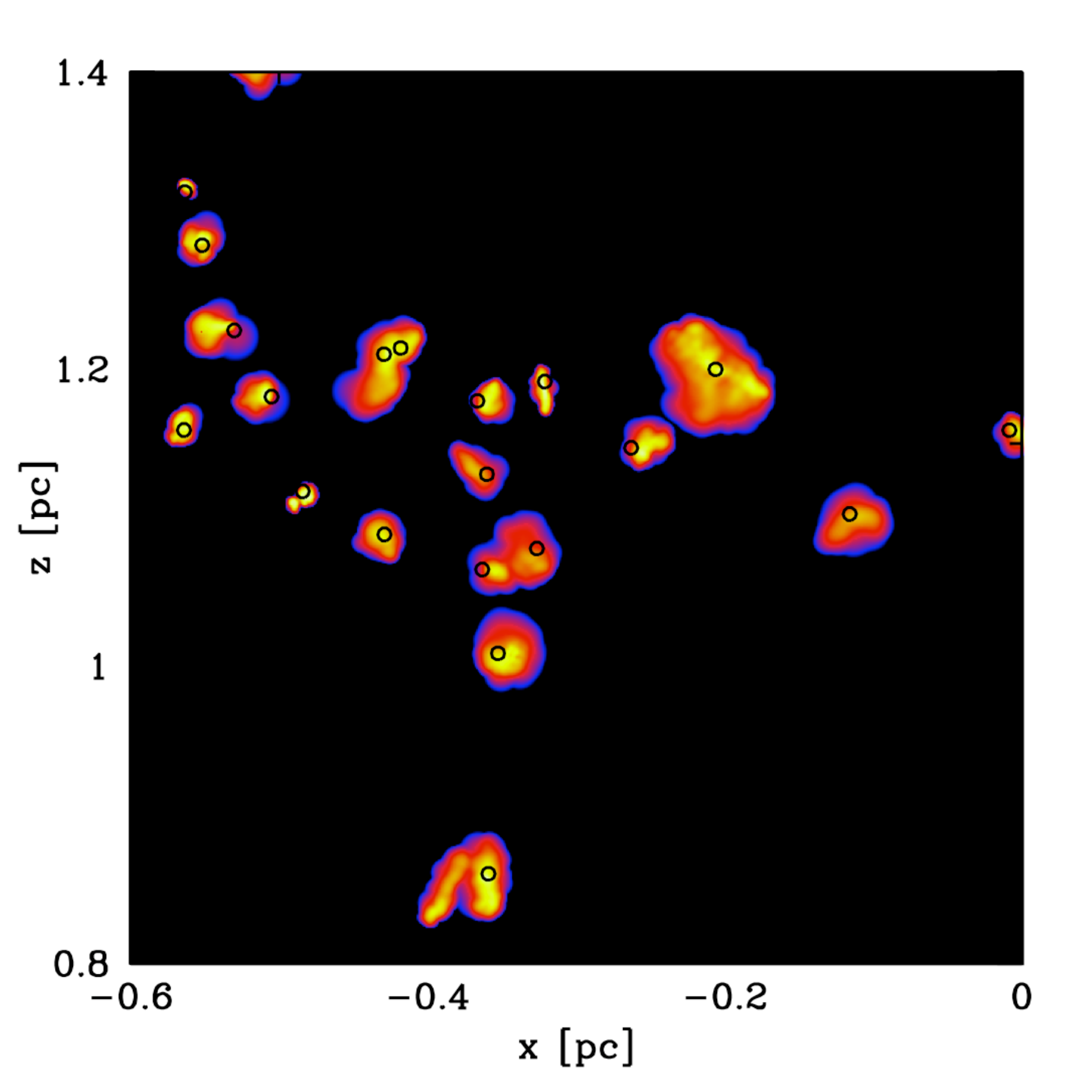} \\
\end{tabular}
\caption{A close up view of some of the pre-stellar pcores identified from Figure \ref{standard}. The left panel shows a zoom of a clustered region in column density and the right panel shows the pre-stellar p-cores identified in it. Colours depict column density and the scale runs from $0.001$ gcm$^{-2}$ (blue) to $10$ gcm$^{-2}$ (yellow). The location of the potential peak of the twenty p-cores in this region are shown by hollow circles. The p-cores are only quasi-spherical and exhibit significant substructure. }
\label{exampleclumps}
\end{center}
\end{figure*}

Figure \ref{exampleclumps} shows a close up view of the p-cores identified at a snapshot in time in one of the clustered regions of our molecular cloud. The p-cores are only quasi-spherical and often are elongated due to the filaments they are formed in. Within the p-cores there is still significant density substructure, and further fragmentation is often observed after the p-core is first bound. Due to this substructure if we had used a traditional clumpfinding algorithm they would have been split into smaller cores. 

\begin{figure}
\begin{center}
\includegraphics[width=2.5in]{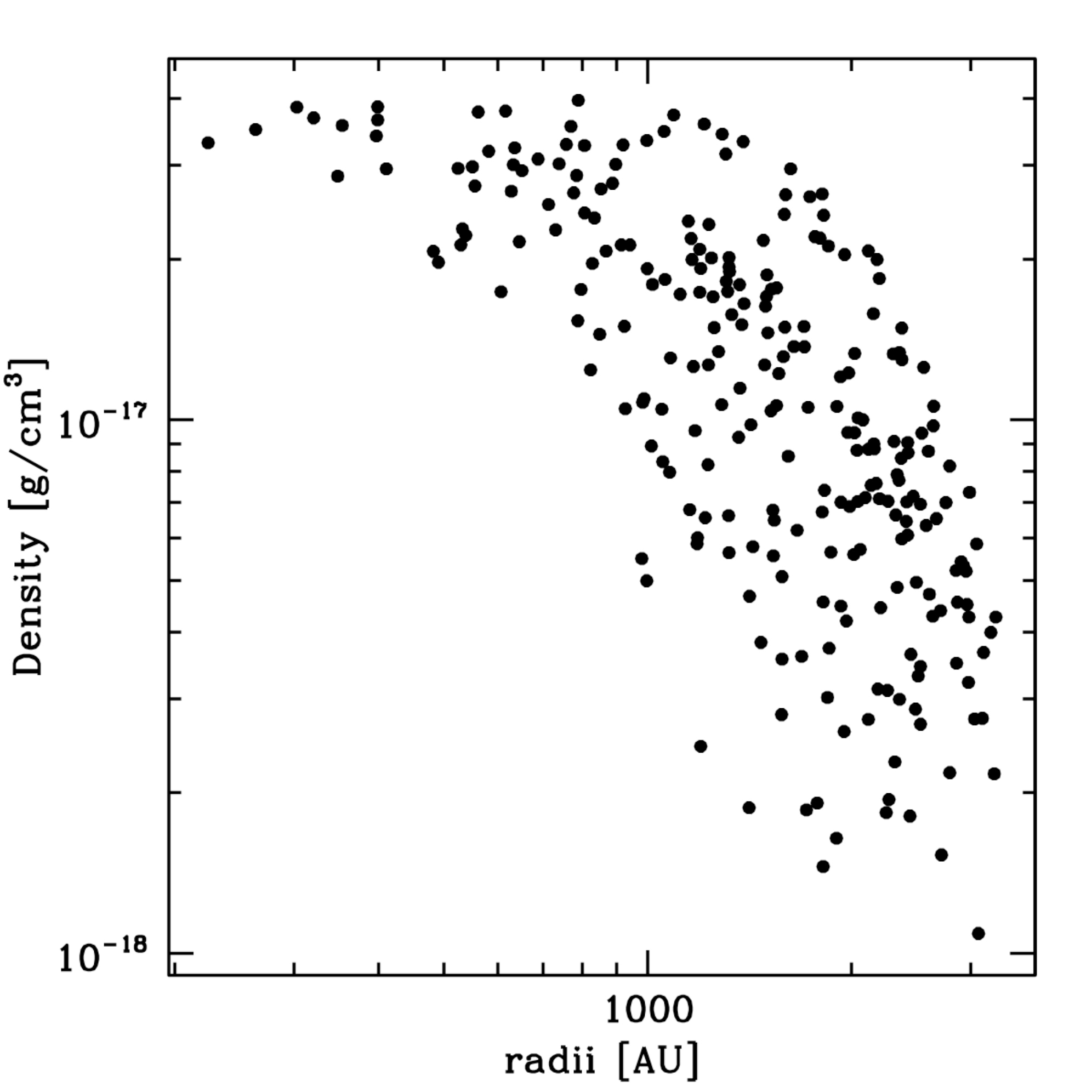}
\caption{The density of the sph particles assigned to a typical p-core plotted radially outward from the peak of gravitational potential. There is considerable dispersion due to substructure, but there is a clear trend showing a flattened central peak and density decreasing outwards.}
\label{denprofile}
\end{center}
\end{figure}

In order to quantify the extent of the p-cores, we use an effective radii, $r_{eff}$, within which 68\% of the mass is contained. As the p-cores are not relaxed, we use the peak in the gravitational potential to define the centre rather than the centre of mass, which produces smoother density profiles. Figure \ref{denprofile} shows the density of the sph particles assigned to a typical p-core, plotted against their distance, $r$, from the central potential peak. The large dispersion shown in the density profile is due to the substructure within the core and the effect of the cloud being embedded in a non-uniform medium.

Nonetheless despite the scatter there is a trend towards a flattened density profile in the central region, as has been seen in observations (e.g \citealp{Ward-Thompson94}). To get an idea of how centrally concentrated the p-cores were, a power law of the form $\rho \propto r^{-n}$  was fitted to the p-core and the best fit value of $n$ determined. We excluded the radius within which the first $10$ \% of the mass was contained from our fit due to the aforementioned flattening.

Due to the dispersion from the density substructure and the core's non-spherical nature, there was a large degree of uncertainty, often as large as $50\%$. Despite this uncertainty, when plotted as a histogram Figure \ref{hprops} (c), we find that the bound cores have exponents clustered around $n=1.36\pm0.35$. This is intermediate between a shallow density increase ($n=1)$ and that of a free-falling envelope ($n=1.5$), and is an indication that our p-cores are still being formed when they become bound. Typically exponents of $n\approx 1.6$ are expected for Class 0 and Class 1 YSO's \citep{Young03} . Conversely, the profiles of the composite p-cores have a wide range of density exponents with an average value of 1.08. 


\subsection{Masses and Sizes}
\label{sec:props}

\begin{table}
	\centering
		\caption{The average clump properties of the p-cores in the bound and composite datasets. R$_{eff}$ is the radius within which 68\% of the mass is contained. The density profile is the best fit value of n for the profile $\rho \propto r^{-n}$}
		\begin{tabular}{l c c }
   	         \hline
	         \hline
	          & Bound & Composite\\
	          \hline
	         Mass (M$_{\odot}$) & $0.70$ & $0.78$\\
	         R$_{eff}$ (AU) & $2.4\times10^{3}$ & $3.7\times10^{3}$ \\
	         $\sigma_{3D}(v)$ $(kms^{-1})$ & $0.27$ & $0.40$\\
		Dynamical Time (yrs) & $2.1\times10^{4}$ & $7.4\times10^{4}$\\
		Density Profile ($n$) & $1.37$ & $1.08$\\
		\hline
		\end{tabular}
	\label{cprops}
\end{table}

\begin{figure*}
\begin{center}
\begin{tabular}{c c c }
\includegraphics[width=2.2in]{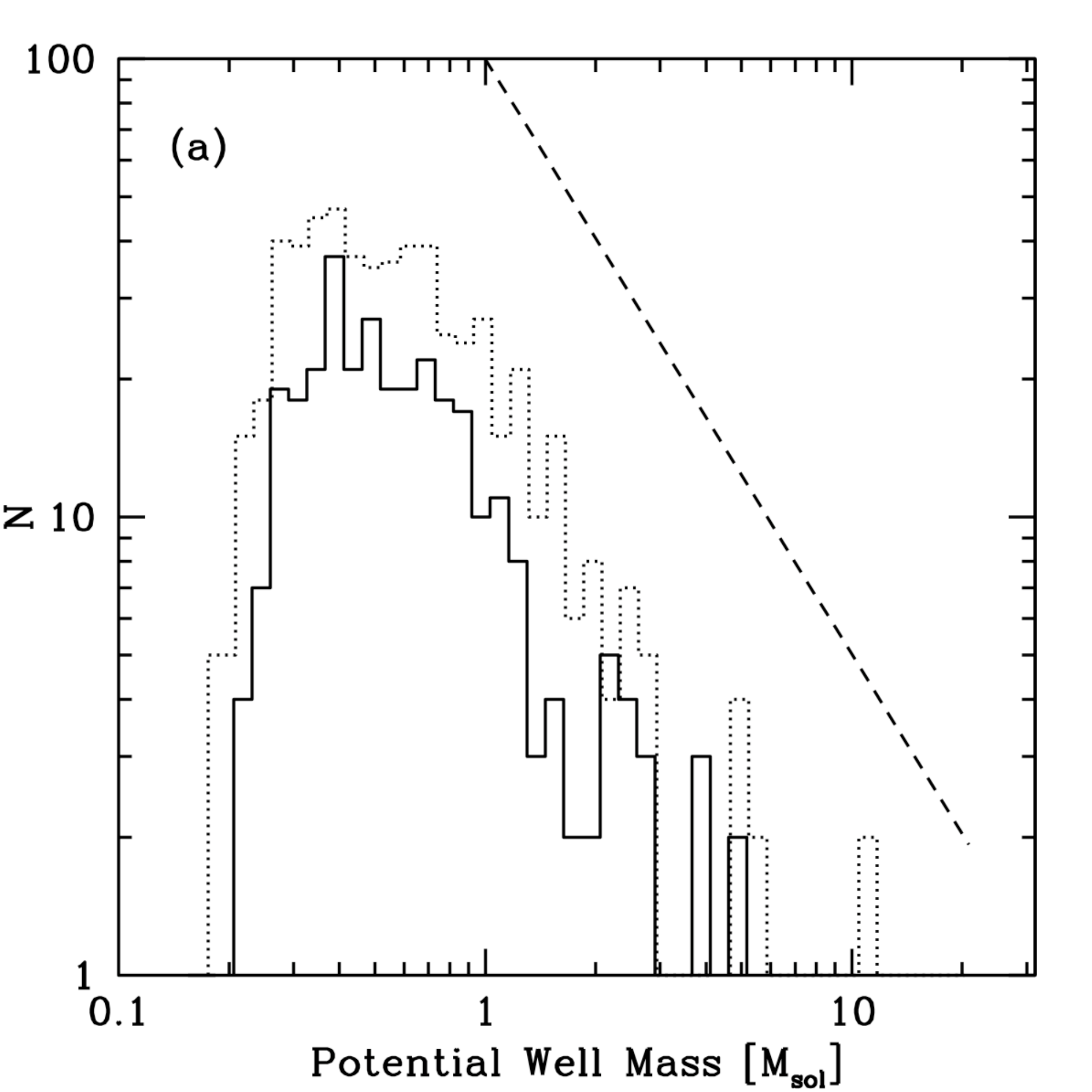} &
\includegraphics[width=2.2in]{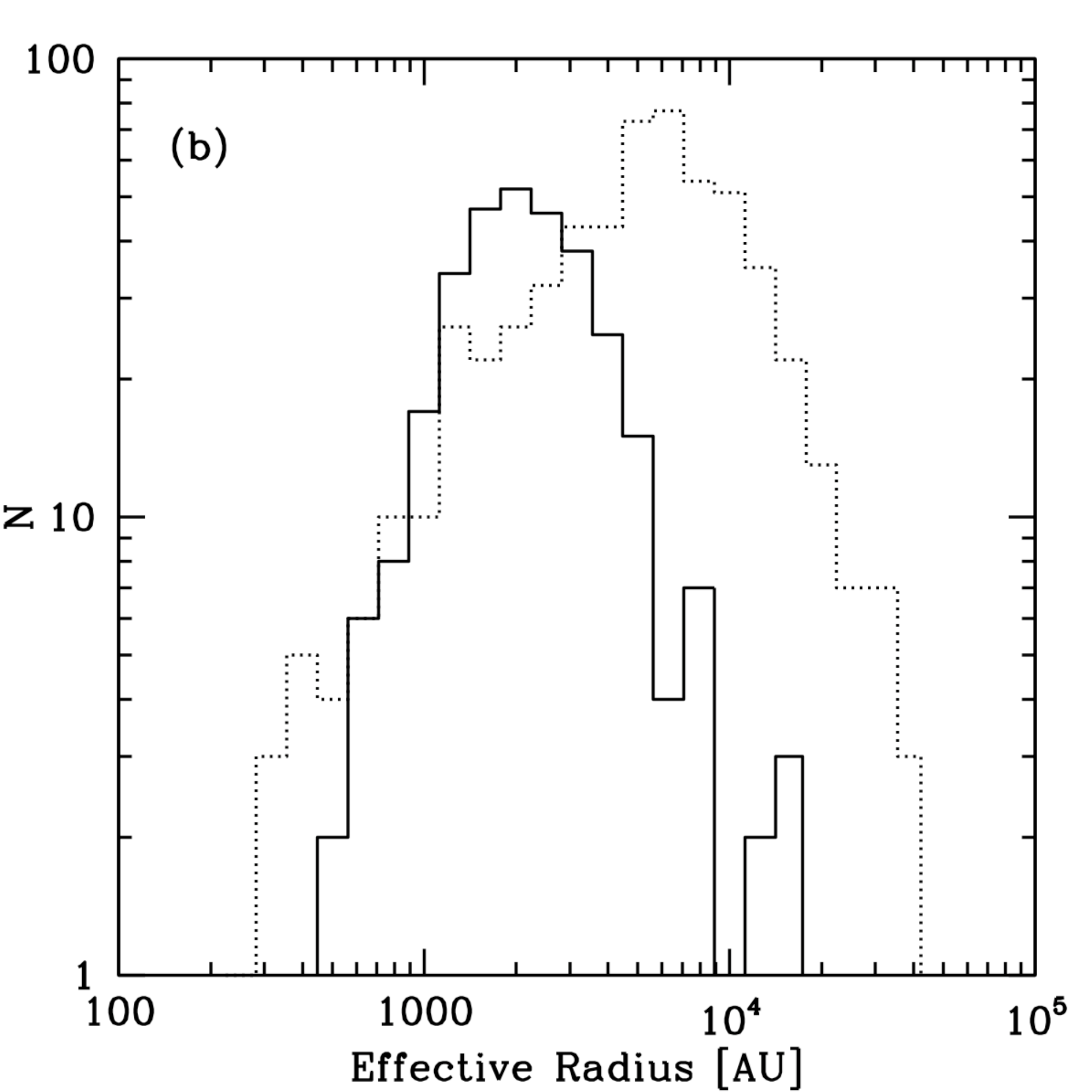} &
\includegraphics[width=2.2in]{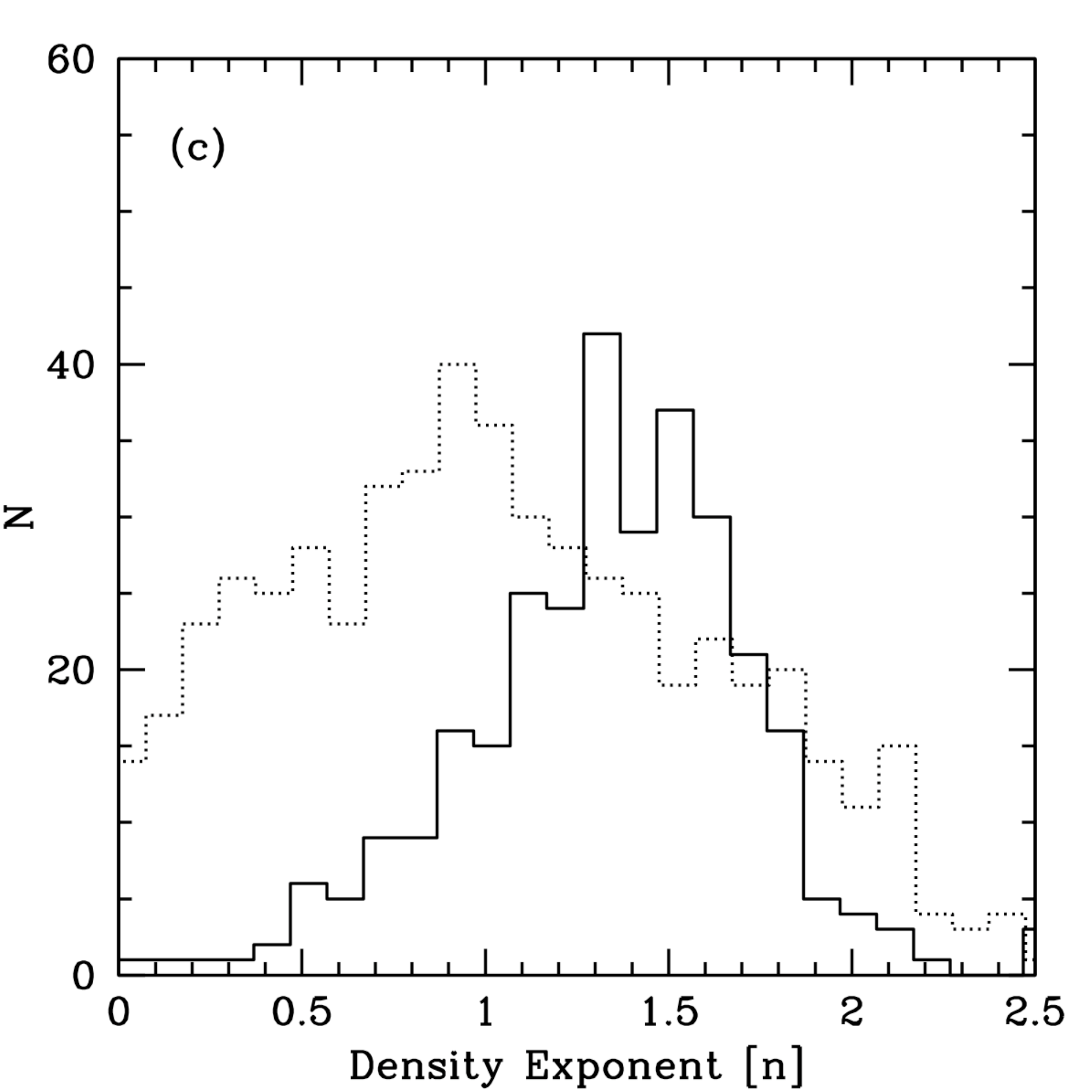} \\
\end{tabular}
\caption{Histograms of the masses \& sizes of the bound (\textit{solid line}) and composite (\textit{dotted line}) p-cores. Panel (a) shows the clump mass function. Masses above $0.2$ M$_{\odot}$ are resolved and the Saltpeter slope is denoted by a dashed line. Panel (b) shows the effective radii. Panel (c) shows the best fit values of $n$ for the profile $\rho \propto r^{-n}$. The p-core mass function resembles the stellar IMF and the p-cores are typically small, centrally concentrated objects.}
\label{hprops}
\end{center}
\end{figure*}

Figure \ref{hprops} (a) shows the clump mass function (CMF) of the bound and composite p-cores, in which there is a clear resemblance to the IMF. Our previous work \citep{Smith08} showed that an IMF-like mass distribution was always obtained from MC structure using the traditional CLUMPFIND algorithm, but it was unclear whether this had any physical meaning. We demonstrate here that the bound cores also follow this distribution. This shows that at some level there is a link between molecular cloud structure, the formation of bound cores, and ultimately star formation. We examine the link between p-cores and stars in Section \ref{sec:masses}

The average p-core masses for the bound and composite data-sets are very similar; $0.70$ and $0.78$ M$_{\odot}$'s respectively. This is broadly consistent with the characteristic stellar mass \citep{Chabrier03}. Both data-sets show this distribution, regardless the fact that members of the composite population are often unbound and do not form stars. 


The distribution of the radii of the p-cores is shown in Figure \ref{hprops}. The bound p-cores have effective radii in the region of $2.4\times10^{3}$ AU, and the composite p-cores have radii of about $3.7\times10^{3}$ AU. Both distributions resemble a lognormal. Unlike the clump mass function, the distribution of the bound and composite data-sets differs in magnitude, which is due to the composite population containing a large number of unbound diffuse clumps.

\subsection{Binding}

\begin{figure*}
\begin{center}
\begin{tabular}{c c c}
\includegraphics[width=2.2in]{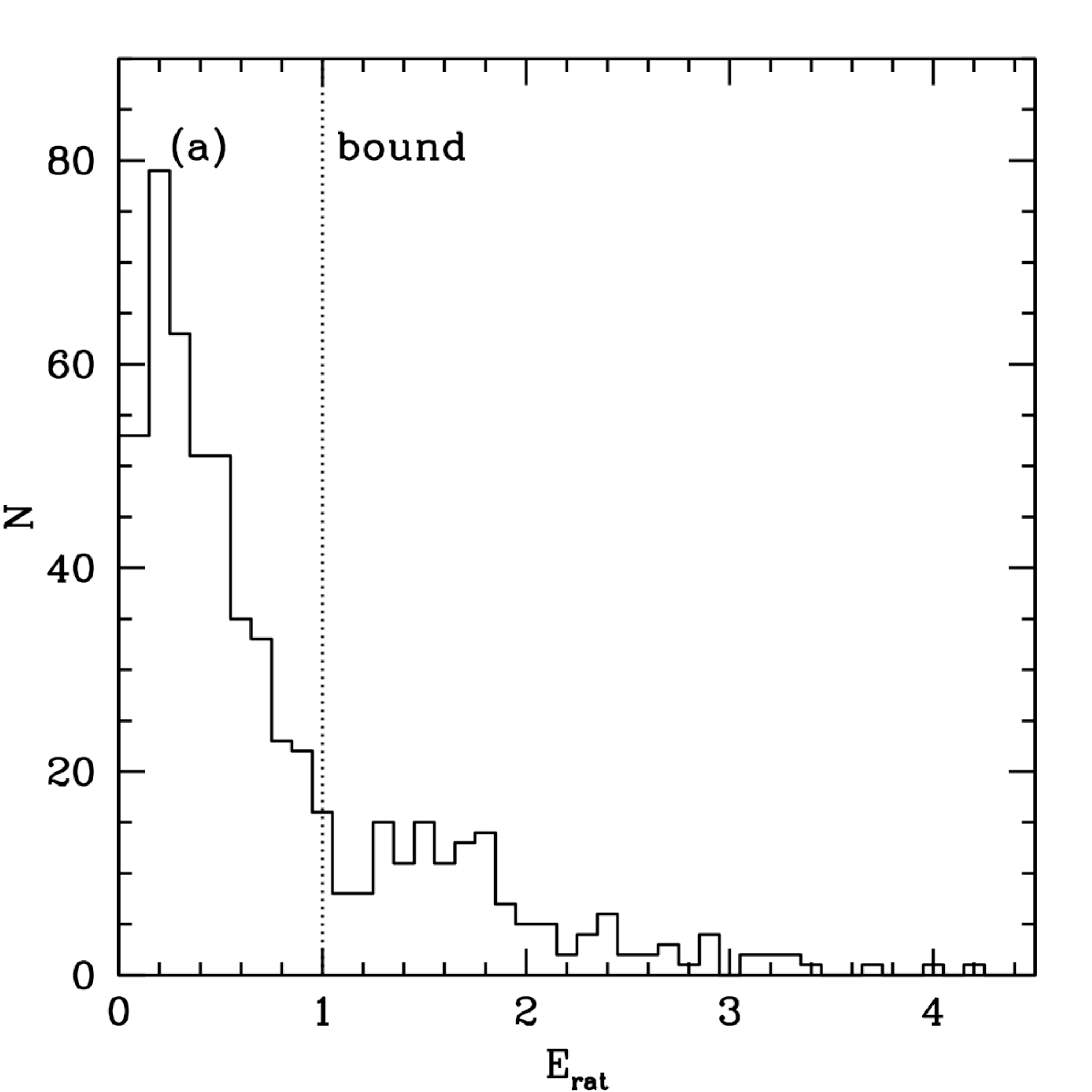} &
\includegraphics[width=2.2in]{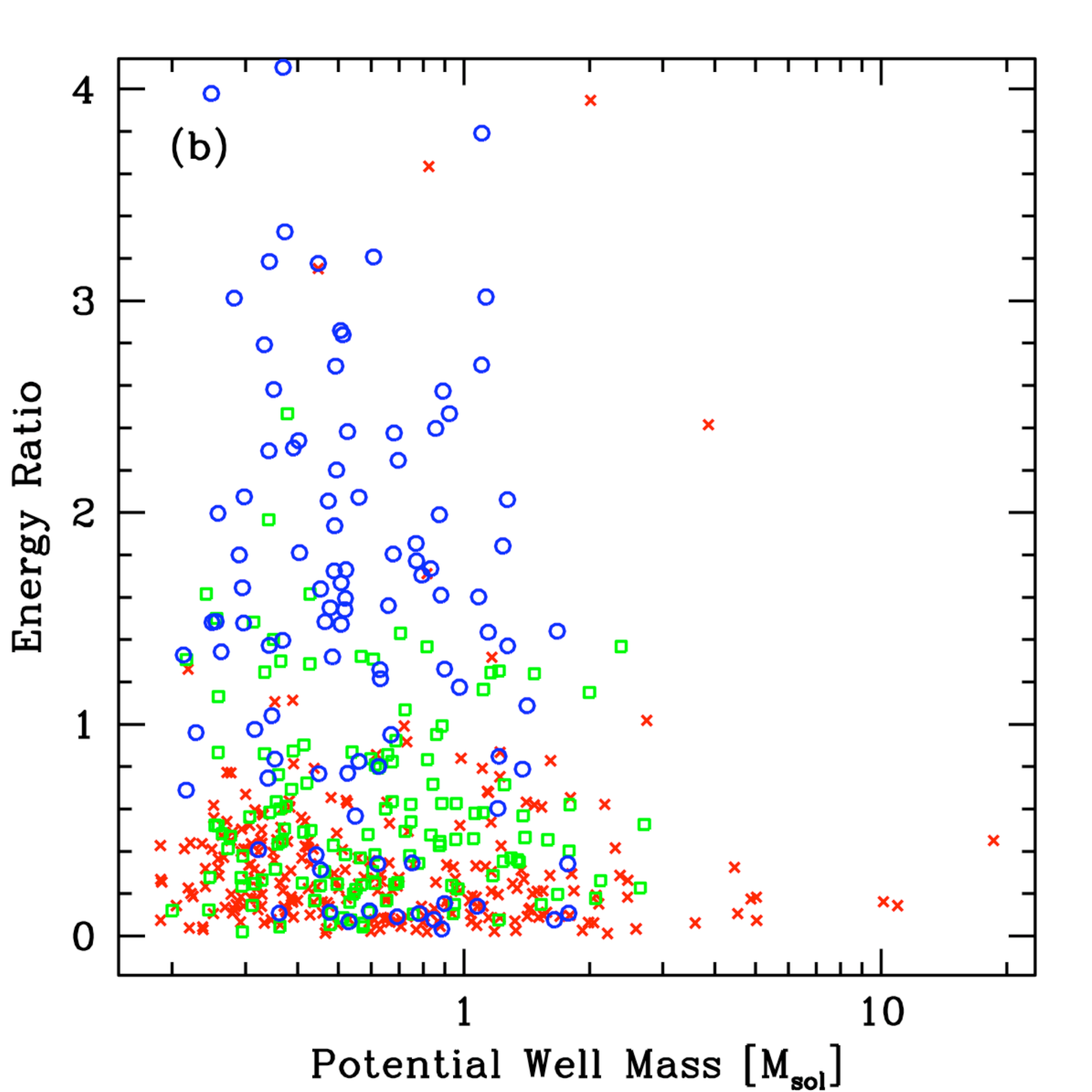} &
\includegraphics[width=2.2in]{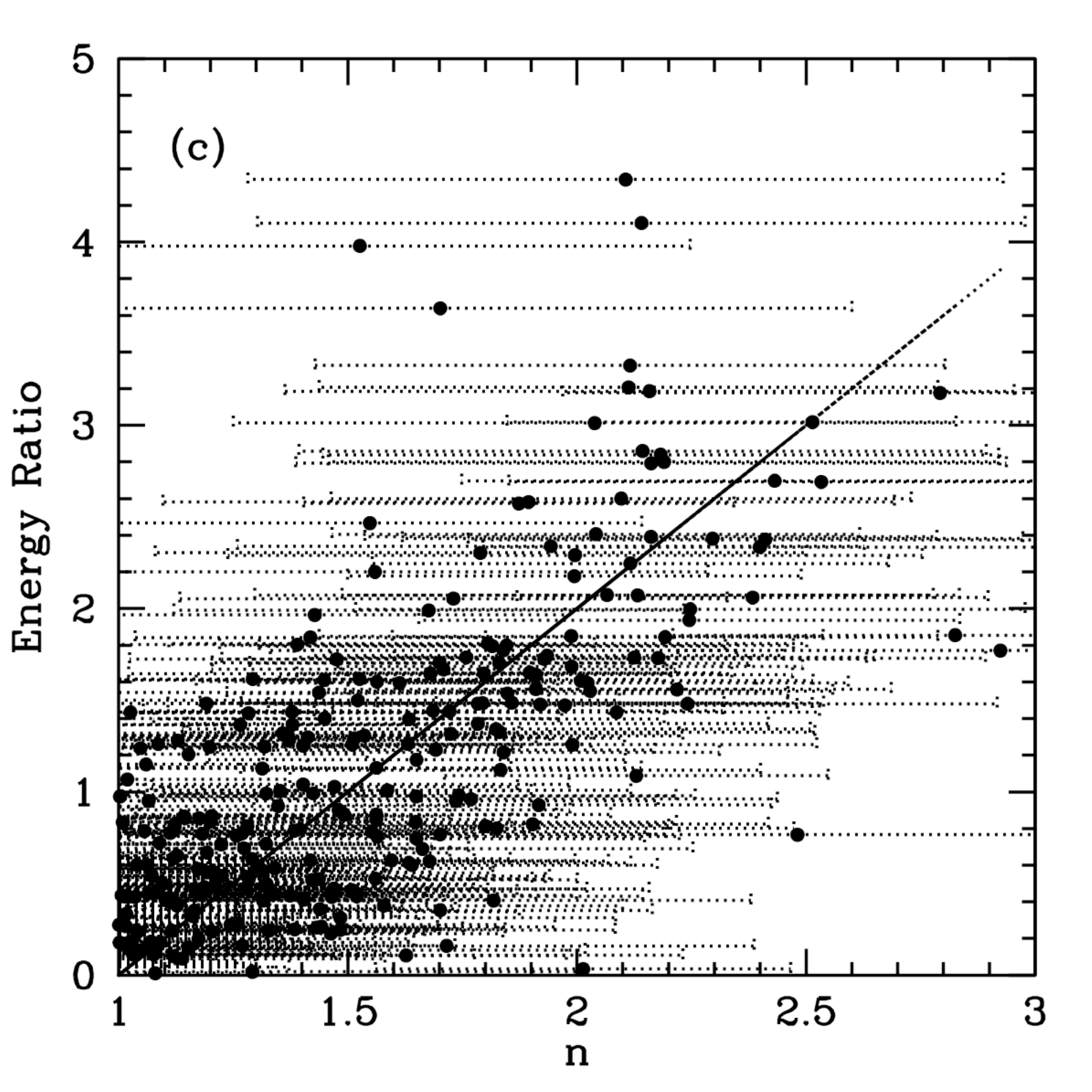}\\
\end{tabular}
\caption{The binding of the composite p-cores. Panel (a) shows a histogram of the energy ratio of the cores, $E_{rat}\ge1$ are bound, where $E_{rat}=|E_{p}| / E_{therm}+E_{k}$. Panel (b) shows the p-core masses plotted against energy ratio, blue circles denote cores with a steep $n>1.5$ density profile, green squares intermediate $1<n<1.5$ profiles, and red crosses shallow profiles. Panel (c) shows the density exponent $n$ plotted against energy ratio with error bars due to the poor fit from density substructure and non-spherical core shapes, the straight line has a gradient of two. There is no correlation between binding and mass, but there is a link to central concentration.}
\label{binding}
\end{center}
\end{figure*}

The binding of the composite data-set is shown in Figure \ref{binding} (a), only $24$\% of the p-cores are bound ($E_{rat}>1$). Moreover, for these p-cores, there is a tail which includes highly bound objects. The composite population consists of p-cores in a wide range of evolutionary states, a large fraction of which will not go on to form stars. Figure \ref{binding} (b) also shows the p-core masses plotted against energy ratio. There is no correlation between binding and mass in our simulation, contrary to some observations where the most massive cores are the most bound (e.g. \citealt{Lada08}). This is due to the tendency of our larger cores to be more diffuse. Due to binding being roughly constant with p-core mass, bound cores appear to be selected with a uniform probability from the composite p-core distribution, which explains the resemblance between the composite and bound CMF's.

The density profile of the mass assigned to the core radially outwards, on the other hand, does show a correlation with binding. In Figure \ref{binding} (b), shallow ($n<1$), intermediate ($1< n <1.5$) and steep ($n>1.5$) density profiles are denoted by circles, squares and crosses respectively. We find that the steepest profiles are nearly all bound, and about half of the intermediate profiles are bound. Figure \ref{binding} (c) shows the density exponents of the p-cores plotted against their energy ratio. When the profiles are shallow, a core is almost never bound, but when the potential well deepens and the core becomes more bound, the density profile steepens as the p-core becomes more centrally concentrated. However, there is a high degree of dispersion in this relation due to the difficulty of obtaining density fits from only quasi-spherical cores which contain substructure. Nonetheless, the steeper density profiles and increased central condensation show the clearest indication of binding from all the observationally visible quanities. This echoes the Bonnor-Ebert \citep{Bonnor56,Ebert55} sphere model often used to model core observations (e.g. \citealp{Johnstone00}), in which collapse begins above a critical ratio of maximum central density to mean density. However our objects are not generally in equilibrium, but are dynamically evolving in a similar manner to that shown in \citet{Ballesteros-Paredes03}.

\begin{figure}
\begin{center}
\includegraphics[width=2.2in]{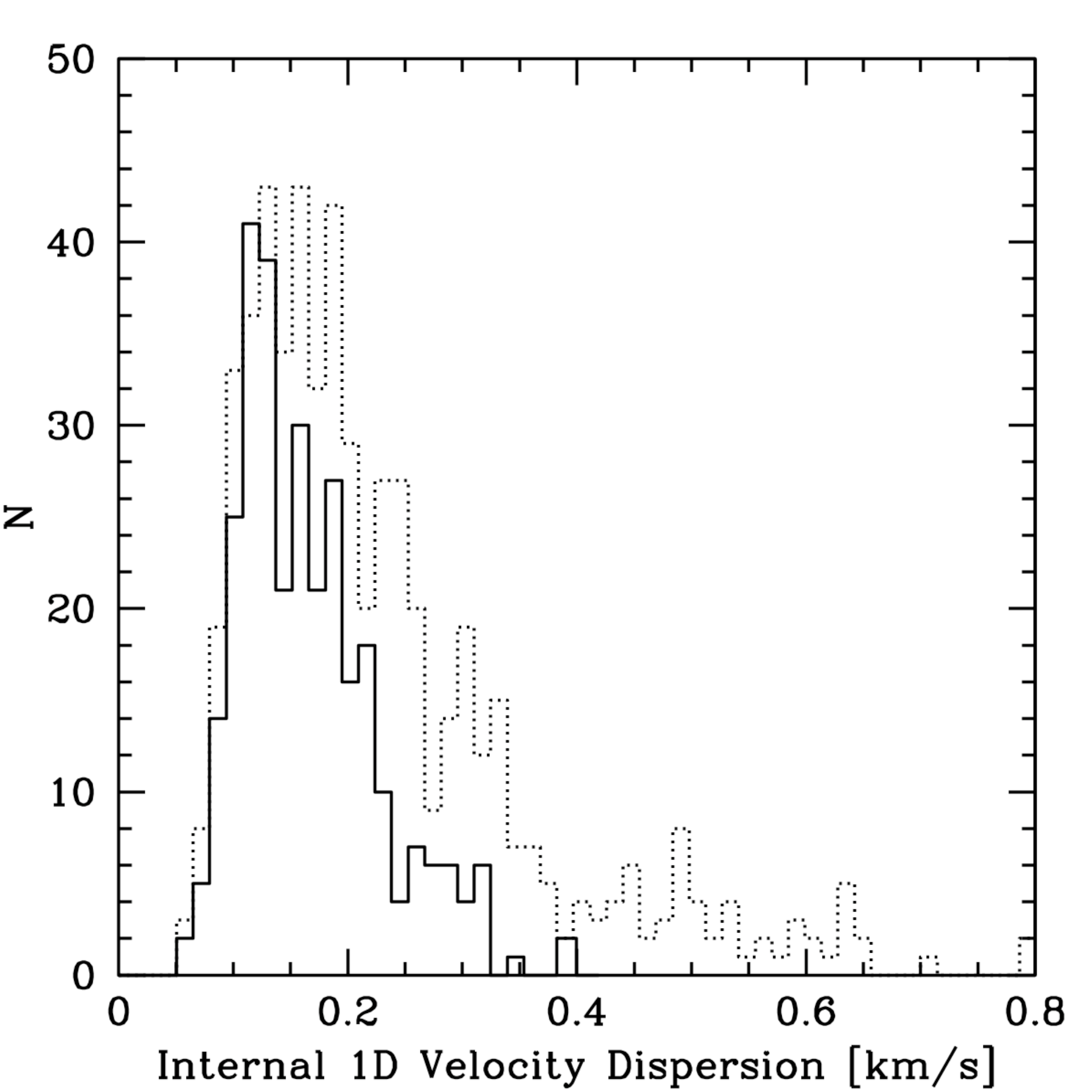} \\
\caption{The one dimensional internal velocity dispersions of the p-cores in the bound(\textit{solid line}) and composite (\textit{dotted line}) datasets. The sound speed of an isothermal gas at 10K is 0.2 kms$^{-1}$ which means our p-cores are generally subsonic}
\label{vdisp}
\end{center}
\end{figure}

Figure \ref{vdisp} shows the internal 1D velocity dispersions of our bound and composite p-cores. Originally 3D velocity dispersions were calculated but these have been converted to the 1D for easier comparison to observations. The mean 3D velocity dispersions were $0.27$ and $0.4$ kms$^{-1}$ respectively, and when converted to a 1D velocity dispersion this becomes $0.16$ and $0.23$ kms$^{-1}$. The sound speed of an isothermal gas at $10$K is $0.2$ kms$^{-1}$, meaning the potential cores are typically just on the verge of becoming supersonic, and are therefore coherent objects \citep{Goodman98}. 

The mean dynamical timescale of a bound cores is calculated using the depth of the potential, rather than the standard practice of using the density, as the gravitational potential is the more clearly defined quantity for our p-cores. We equate the potential and kinetic energies using the virial theorem to obtain a typical velocity which is used to find the dynamical time as shown below, where $R_{eff}$ is the effective radius of the core and $\phi$ the gravitational potential.
\begin{equation}
\label{tdyn}
t_{dyn}=\frac{R_{eff}}{\sqrt{\phi}}
\end{equation}
The average dynamical time is $2.8\times10^{4}$ yrs, and for the composite data-set it is about two times larger. This is more than ten times shorter than the dynamical time of the molecular cloud as a whole; meaning that several generations of cores can form and evolve throughout the lifetime of the simulated molecular cloud.

\subsection{ The Core Mass Function with Time}
\label{sec:imf}

\begin{figure}
\begin{center}
\includegraphics[width=2.2in]{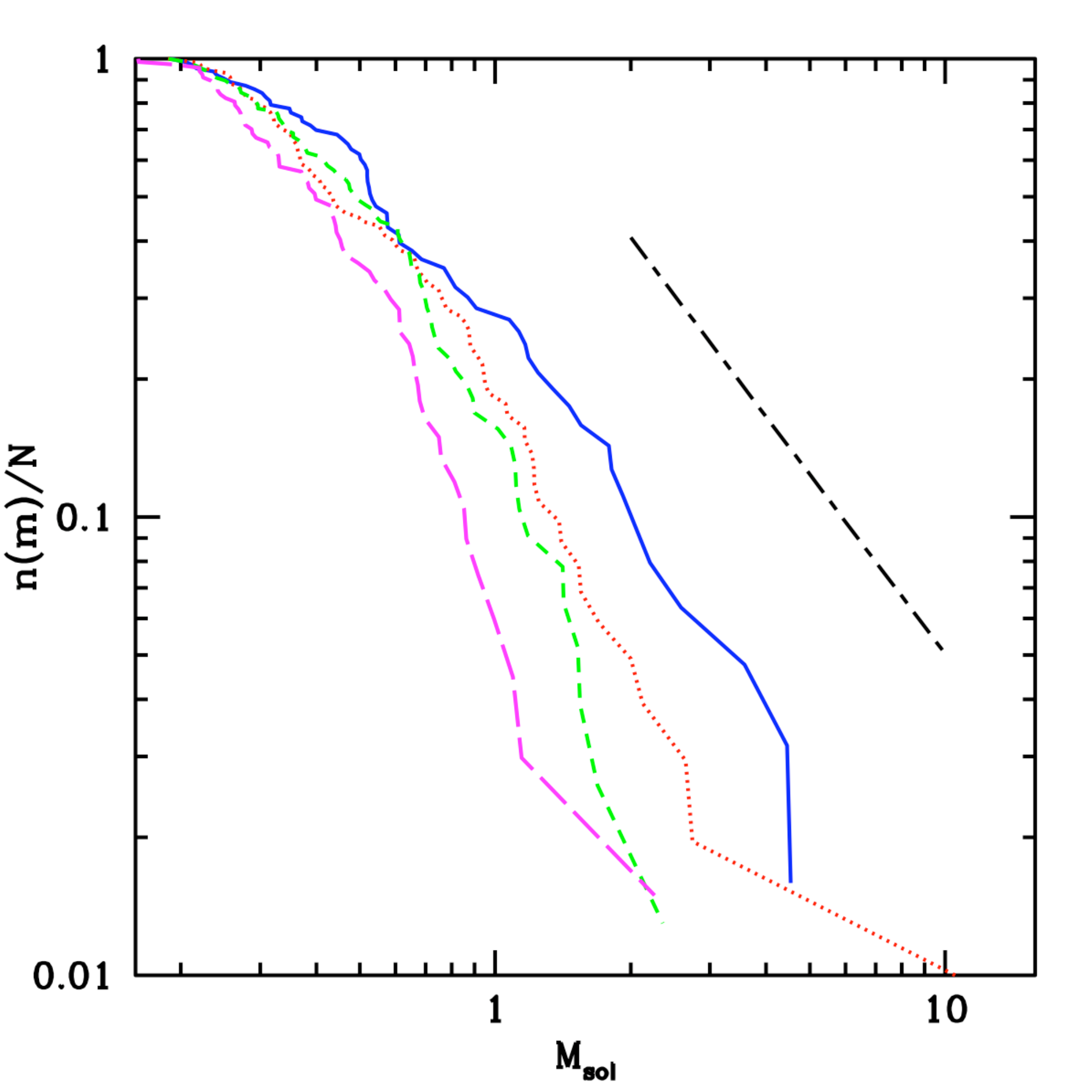}
\caption{The cumulative mass functions from snapshots at \textit{solid} 0.6 $t_{dyn}$, \textit{dotted} 0.8 $t_{dyn}$,\textit{short dashed} 1 $t_{dyn}$ and \textit{long dashed} 1.2 $t_{dyn}$. The dot-dashed line shows the Salpeter slope. The mass function gets steeper with time as the high mass p-cores are formed earlier.}
\label{cmf_cst}
\end{center}
\end{figure}

Figure \ref{cmf_cst} shows the mass functions of the pre-stellar (without sinks) p-core snapshots before they were merged into the composite data-set to integrate out time effects. Snapshots are shown at $0.6$ to $1.2$ simulation dynamical times ($t_{dyn}=4.7\times 10^{5}$ yrs) at intervals of $0.2$ $t_{dyn}$ . 
The high mass end of the clump mass function gets steeper with time, meaning that more massive cores were formed at the beginning of the simulation than at the end. This is probably due to cores at later times being more likely to form in the vicinity of an existing potential well. This will tidally truncate the size of the core, and hence limit the mass it can accrete. In our simulation the major regions of star formation are formed at the same time, and hence their evolution is artificially synchronous. If this were not the case the steepening of the mass function would not have been visible. We aim to further discuss the effect of time on the mass function in a future paper. 

\section{Clump Masses \& Stellar Masses}
\label{sec:masses}

Since the first observations of dense cores, a direct causal relation between them and stars was proposed (e.g. \citealp{Motte98}).  \citet{Alves07} go further, and find that core masses and stellar masses are related with an efficiency of $0.3$. However, this neglects the influence of environmental factors on the core during the accretion process. Moreover, \citet{Swift08} have shown that cores can have variable efficiencies or form multiple stars and still generate the expected IMF.

\begin{figure}
\begin{center}
\includegraphics[width=2.2in]{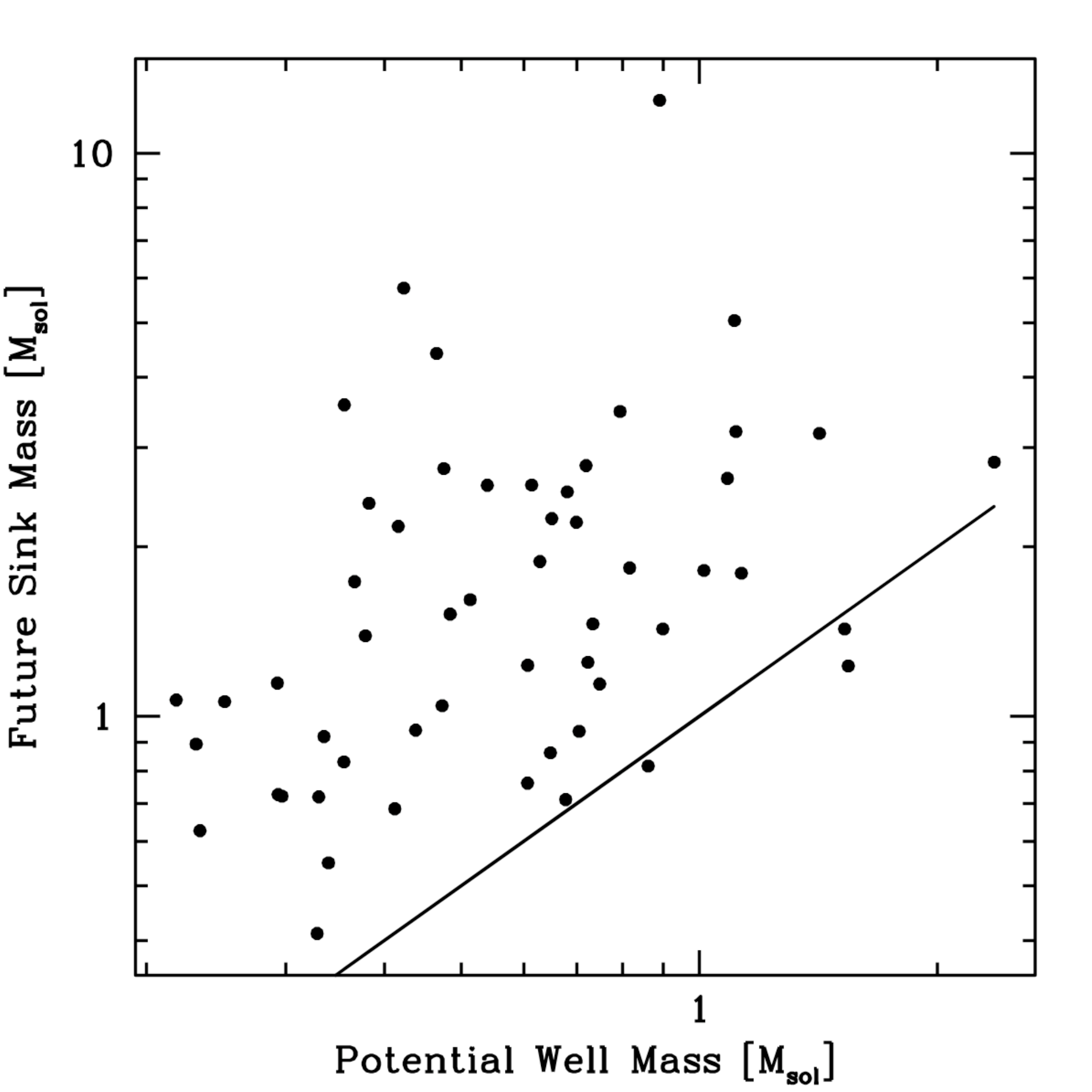}
\caption{The connection between p-core masses at a snapshot in time and their sink mass when the simulation was stopped. The solid line shows a 1-1 correspondence. There is a poor correlation between p-core mass and the total sink mass formed from them.}
\label{corrtfixed}
\end{center}
\end{figure}

We now examine the correlation between core masses and their resulting stellar masses. To investigate how much of the p-core mass is available to form stars, we trace the mass of the sink particles formed from them with time. If a p-core forms more than one sink, we add their masses. This means we are actually tracing the correspondence between p-cores and the stellar systems formed from them. Figure \ref{corrtfixed} shows the masses of the sink particles formed as a function of their p-core masses evaluated at a single snapshot in time. Sink masses are recorded at the end of the simulation and the p-cores in the snapshot population that did not form sinks are neglected.

There is no clear relation between the p-core masses and the sink masses. The sink masses all lie near  or well above the p-core mass values showing the importance of subsequent accretion. Note that as feedback is not included and all sinks that form from a core are counted, it would be surprising if the efficiency of sink formation from cores was much less than $100\%$. Instead it is more than a factor of two higher.  The actual stellar mass formed would of course be less than the sink mass, which simply represents the mass which would reach the inner disk of an accreting proto-star. During the accretion process mass will be ejected via a jet, a process which is not modelled here. However, if there is no correlation between the mass reaching the forming proto-star there is unlikely to be a correlation with the final star either. In fact the correlation is most likely to be further worsened by this process. For instance \citet{Matzner00} find a star formation efficiency of between $25\%-70\%$ in model cores with feedback, depending on geometry. Interestingly, they also find that the stellar IMF formed from the cores is insensitive to core efficiency. Since all observations are necessarily at a snapshot in time this work suggests that the final stellar mass cannot be predicted accurately from the observed core mass even if our p-cores were directly observable. 

\begin{figure*}
\begin{center}
\begin{tabular}{c c}
\includegraphics[width=2.5in]{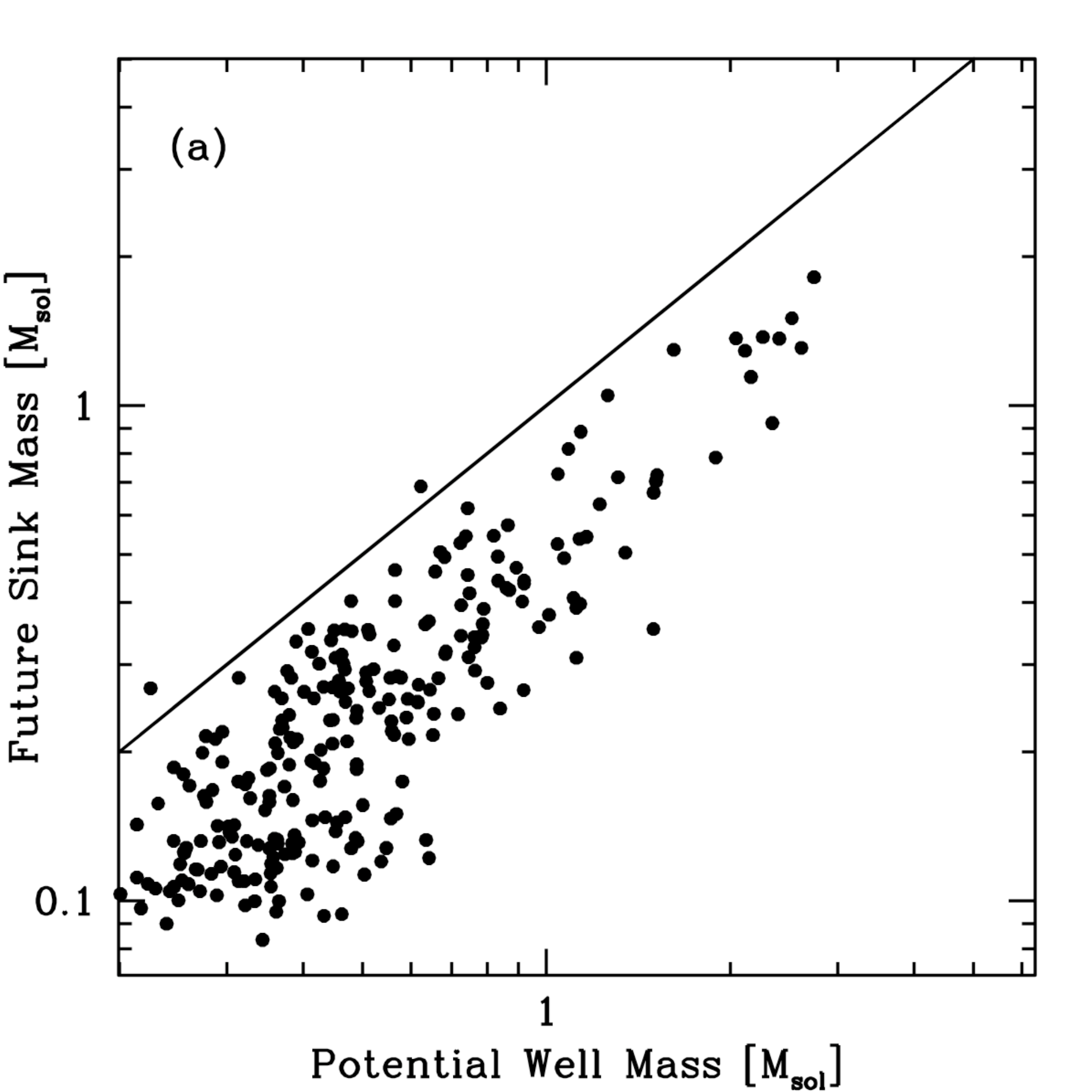}
\includegraphics[width=2.5in]{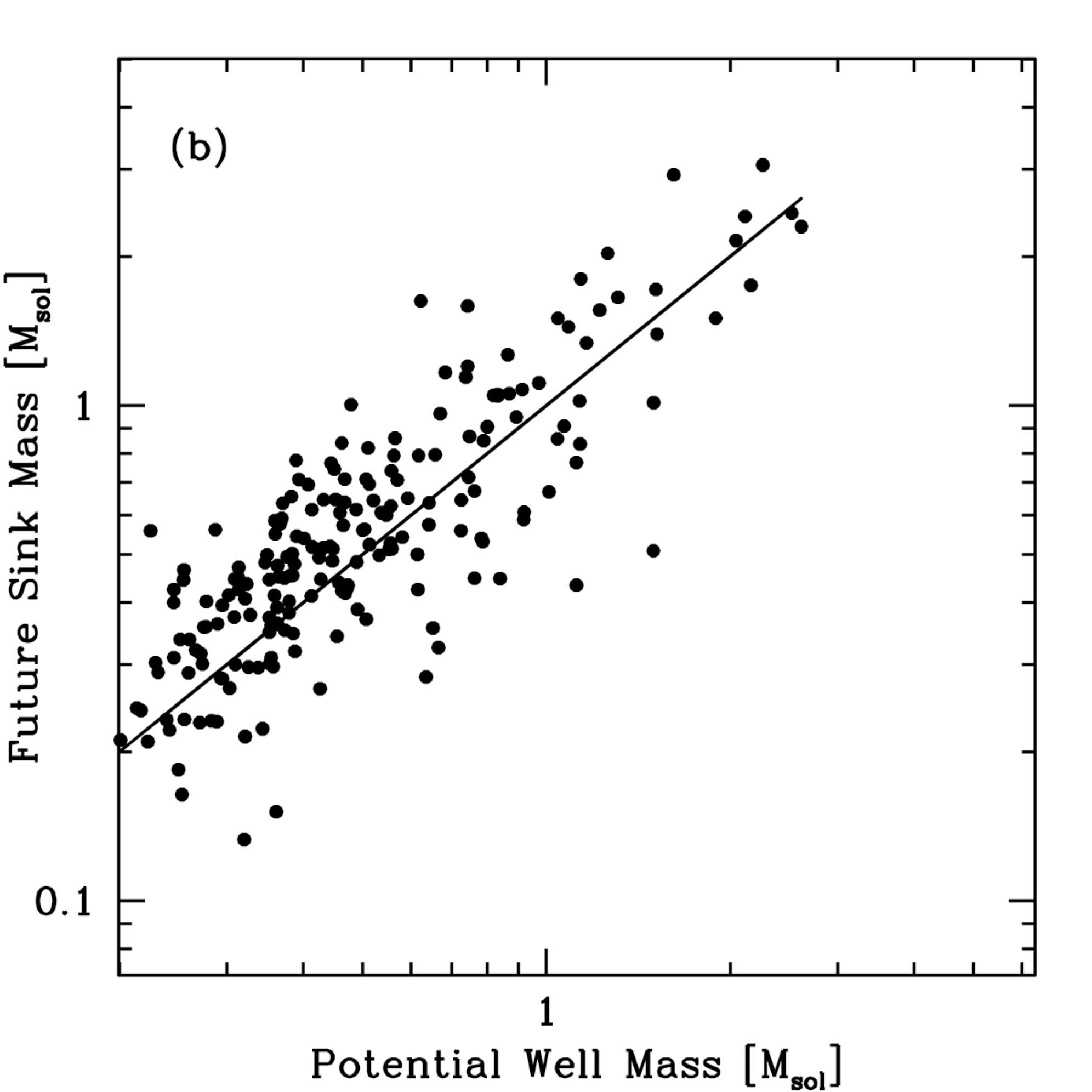}\\
\includegraphics[width=2.5in]{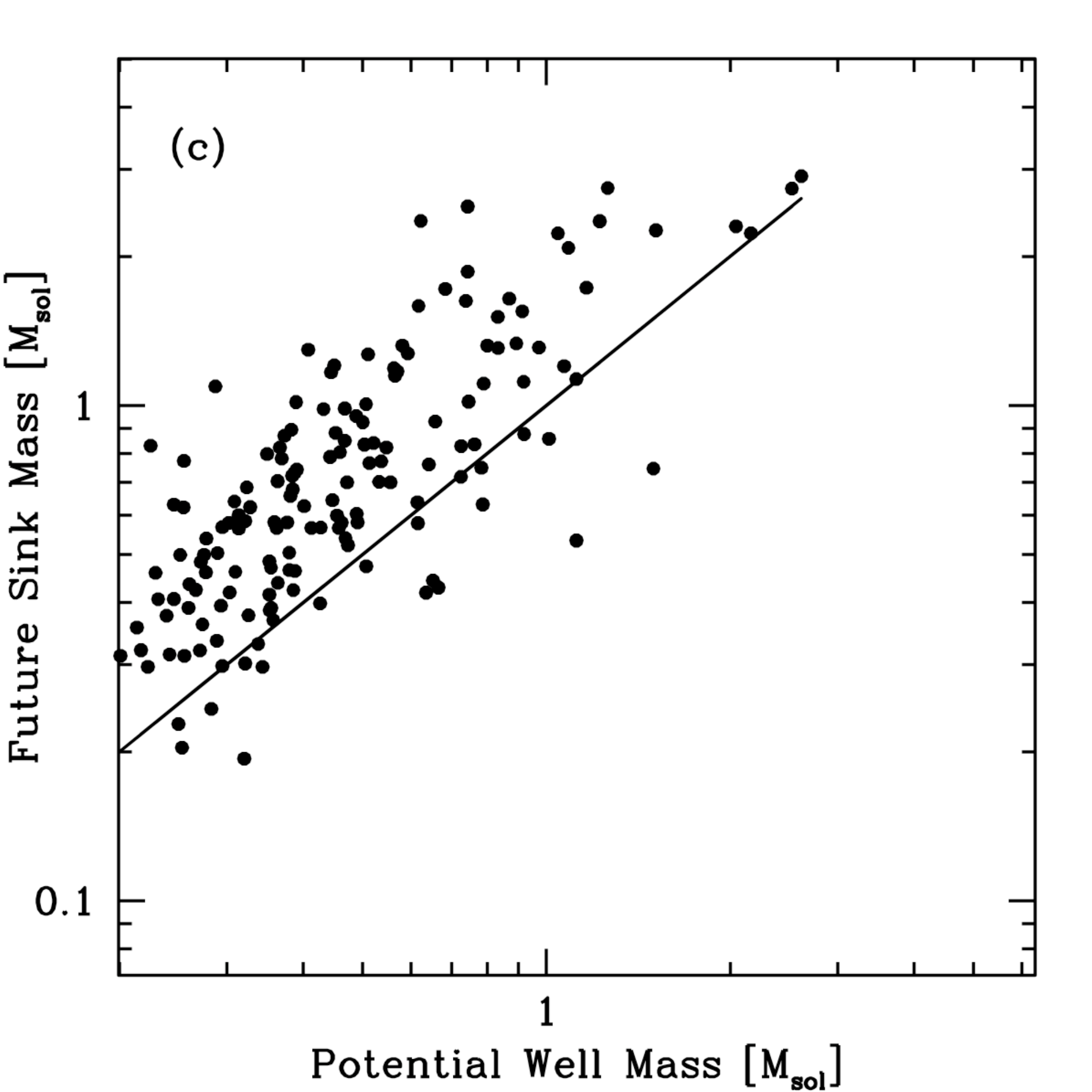}
\includegraphics[width=2.5in]{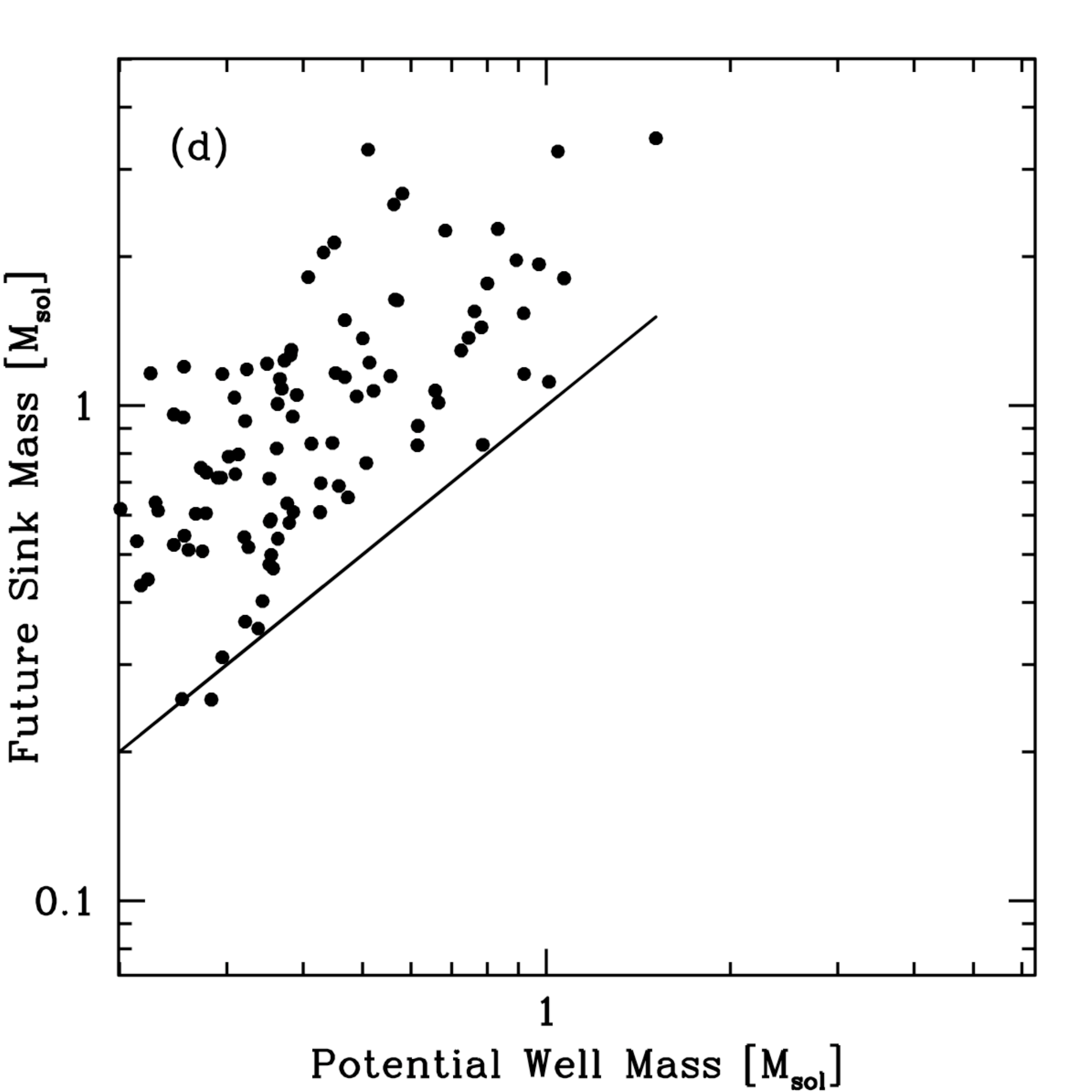}
\end{tabular}
\caption{The connection between clump mass and sink mass at successive dynamical times. Panel (a) $t_{dyn}=1$, (b) $t_{dyn}=2$, (c) $t_{dyn}=3$, (d) $t_{dyn}=5$. The solid line shows a 1-1 correspondence. There is now a clear connection between p-core mass and sink mass, but it still shows significant dispersion.}
\label{correspondence}
\end{center}
\end{figure*}

One complication of Figure \ref{corrtfixed} is that the cores are all at different evolutionary states. We can address this by using their properties at the point in time where each is first bound, as this is when fragmentation and collapse begins, and when the best correlation to the p-core mass would be expected. We record the total mass in sinks formed from these p-cores at successive core dynamical times after they are first bound. In our simulation there is no mechanism to halt accretion, but as the p-core dynamical times ( Equation \ref{tdyn}) are short, we should be finding the sink masses before this is an issue. Figure \ref{correspondence} shows the masses of the sinks after $1$, $2$, $3$ \& $5$ dynamical times. 

When comparing the sink mass at equal evolutionary stages of the bound p-cores, the correspondence is a lot tighter than that in Figure \ref{corrtfixed}. At $1$ dynamical time after the p-core is bound the sinks are below the 1-1 correspondence line. At $2$ dynamical times they are around this level. These two stages, therefore, follow the accretion of the initial bound core, and so it would be surprising if this correlation did not exist! However, there is considerable scatter in this trend, with sinks masses being as much as a factor of two away from the trend in either direction. This indicates that some cores are accreting their envelope at a faster rate in these initial stages.

As suspected, therefore, there must be a direct connection between bound cores and star formation. This improved correlation of p-core mass and sink mass is found at a specific point in time, when first bound.  This leads us back to the original scheme of star formation proposed by \citet{Jeans1902}, as our potential bound cores represent the local Jeans mass at the point of collapse.

 At $3$ and $5$ dynamical times the cores are accreting from their wider environment i.e. material which became bound to the cores subsequent to when $E_{rat}>1$ for the first time. The correlation between core masses and sink masses becomes increasingly dispersed in log-space as accretion wipes out the direct correspondence and increases the offset as the sink-masses grow with time. Cores surrounded by a plentiful reservoir of gas on all sides will be more successful at these stages than those in a narrow filament for example.

\begin{figure}
\begin{center}
\includegraphics[width=2.5in]{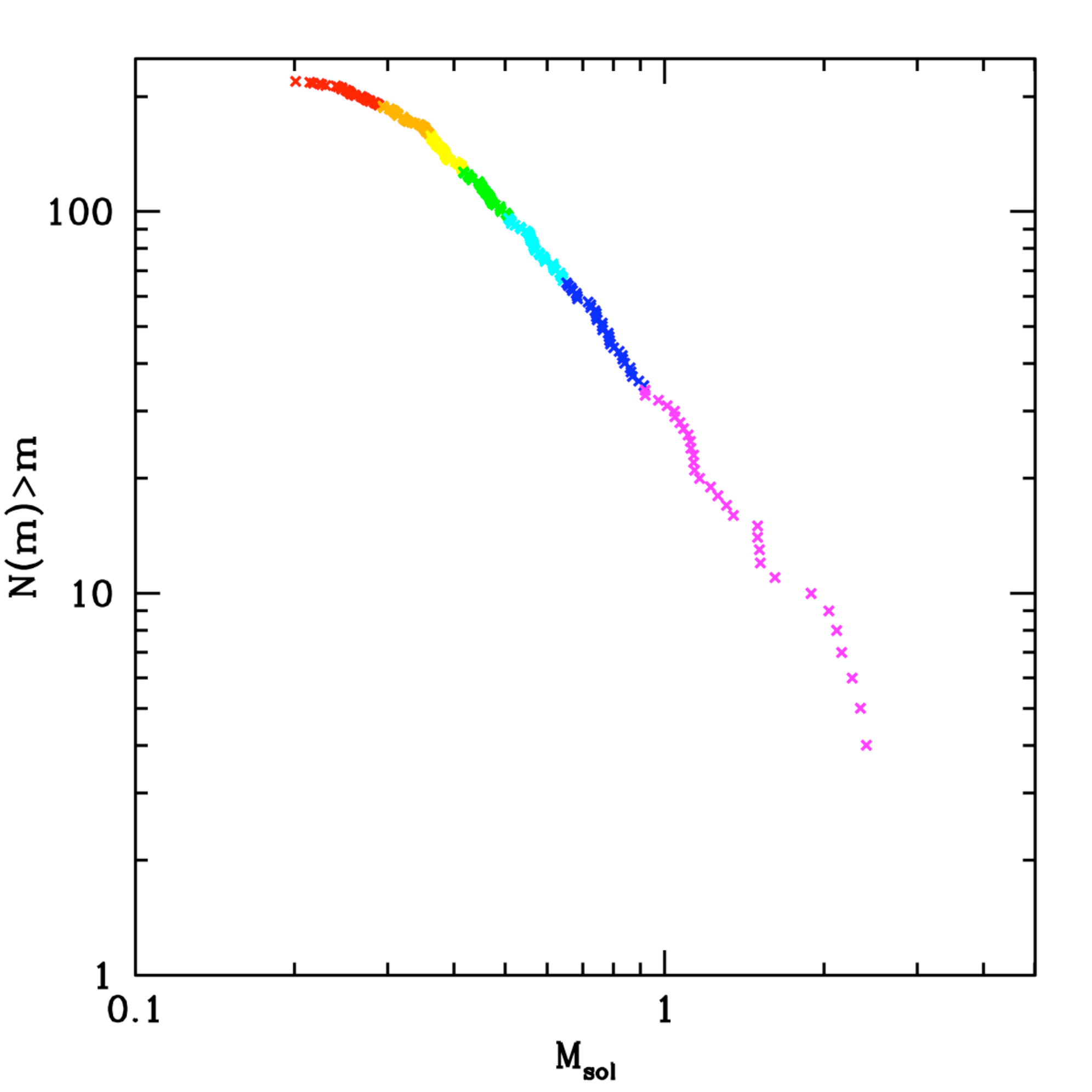}
\caption{The CMF of the p-cores with mass bins denoted by different colours.}
\label{rainbowCMF}
\end{center}
\end{figure}

\begin{figure*}
\begin{center}
\begin{tabular}{c c}
\includegraphics[width=2.5in]{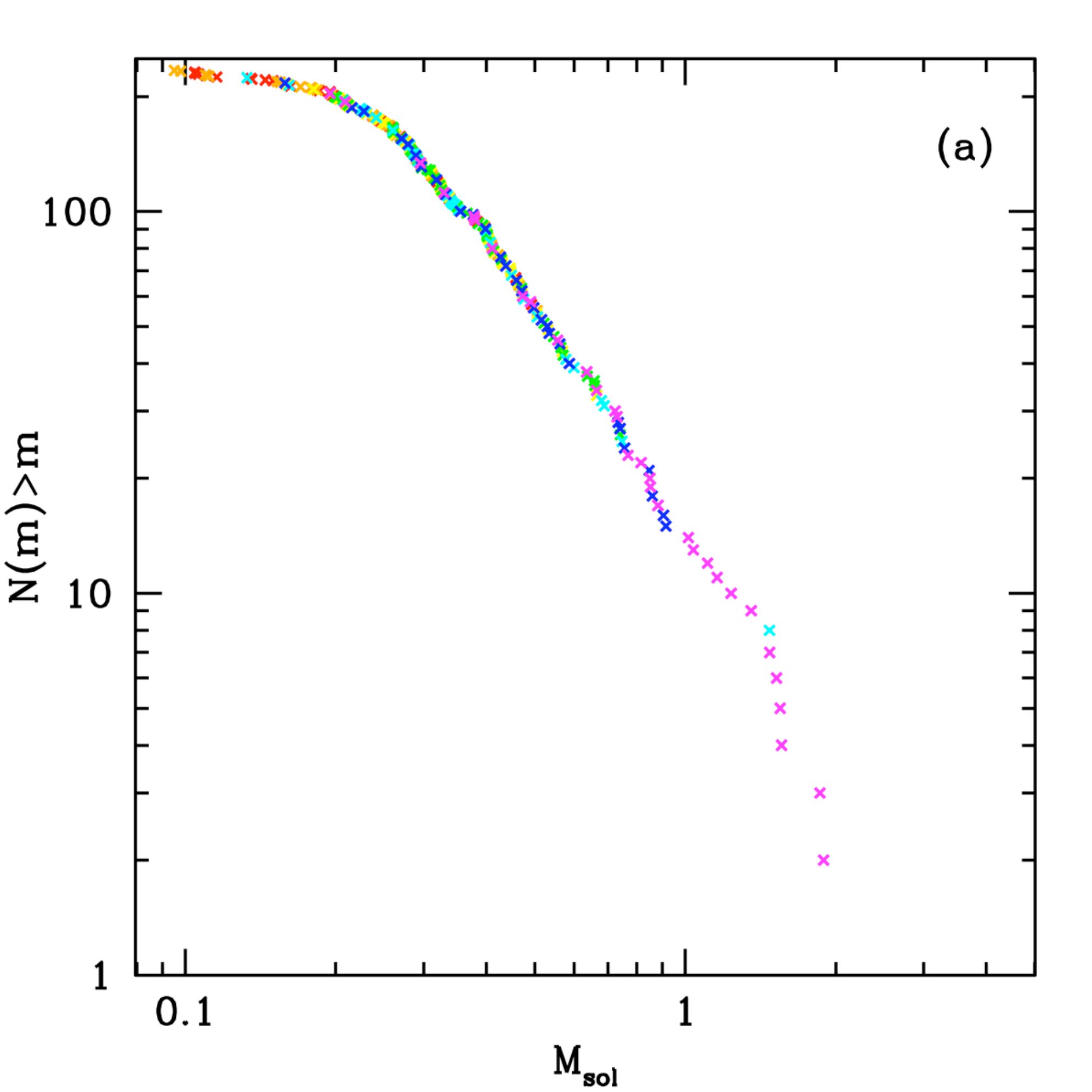}
\includegraphics[width=2.5in]{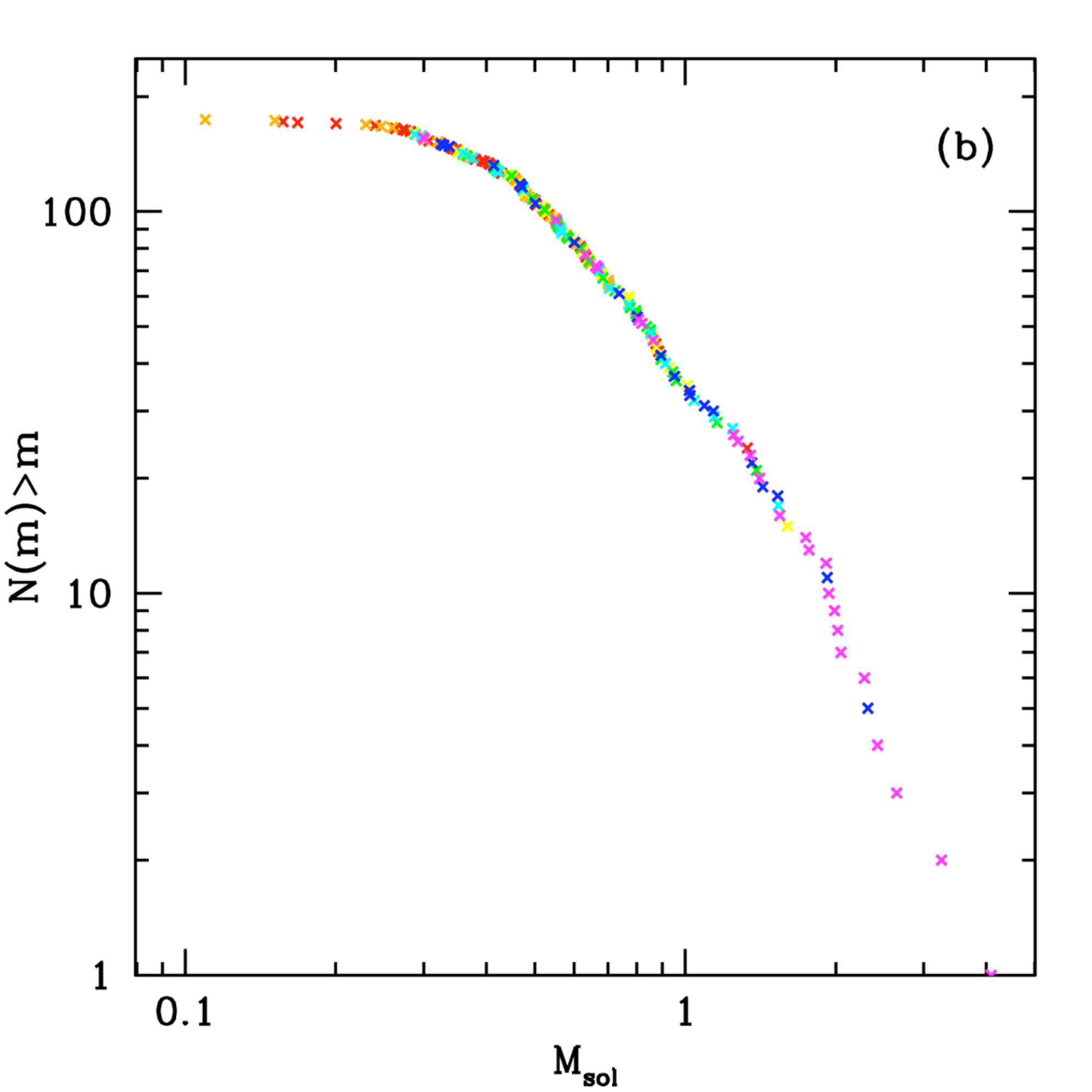}\\
\includegraphics[width=2.5in]{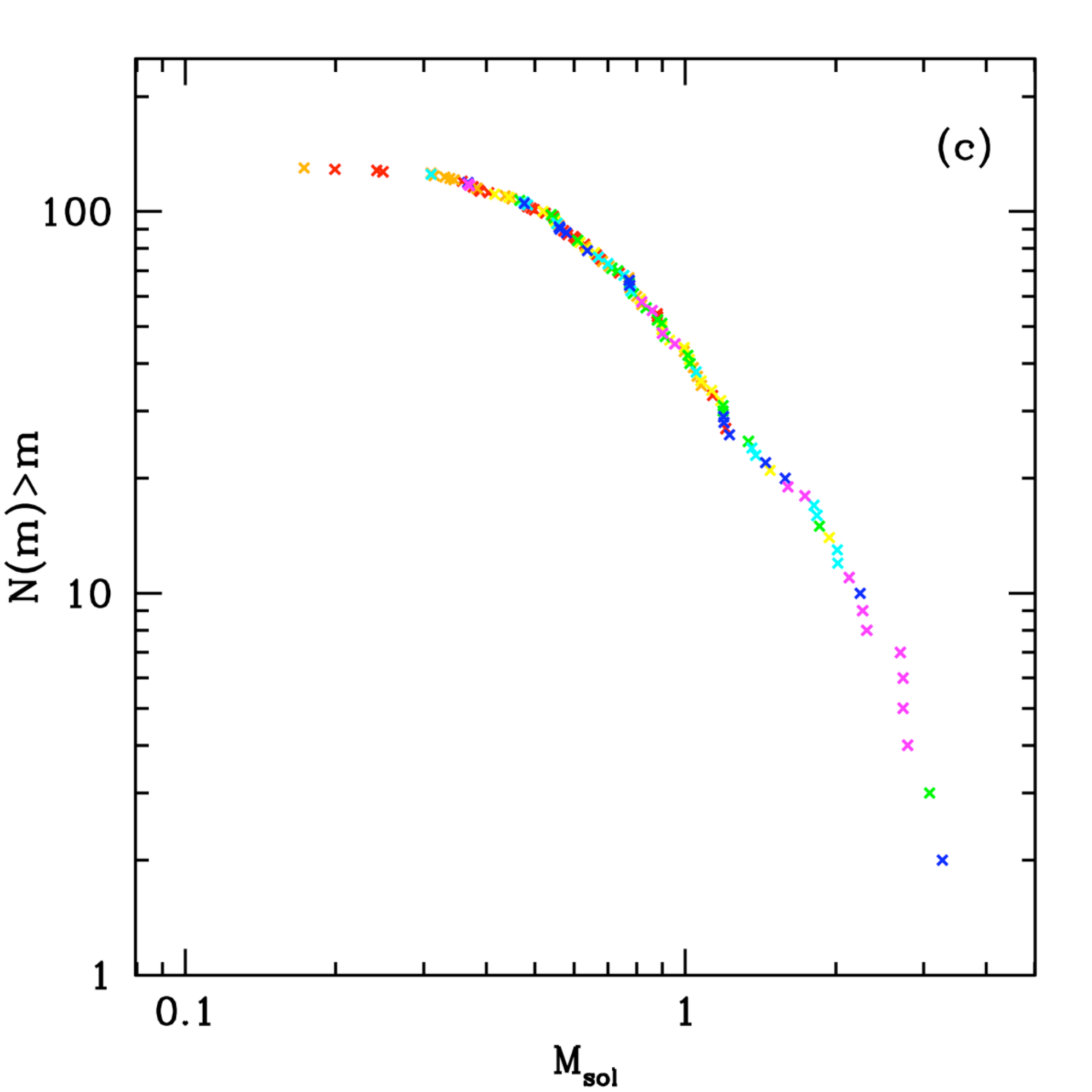}
\includegraphics[width=2.5in]{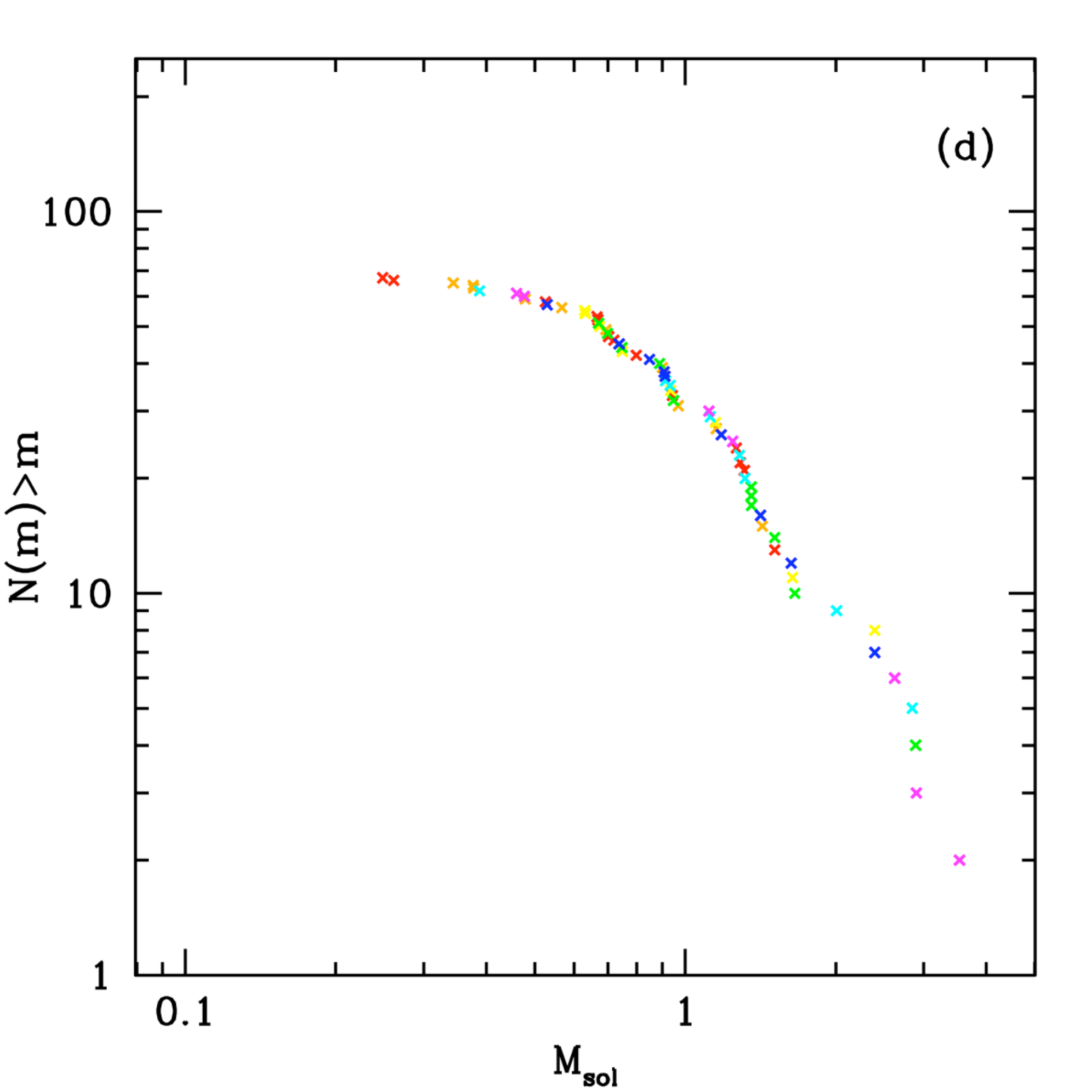}\\
\end{tabular}
\caption{The initial mass functions of the sinks formed from bound cores the colours are the same as their parent p-cores shown in Figure \ref{rainbowCMF}. The IMF's are recorded at intervals of $1,2,3$ \& $5$ p-core dynamical times. The mass bins get mixed up with time, but the shape of the mass function always resembles the stellar mass function despite the inexact correspondence between core mass and sink mass.}
\label{rainbow}
\end{center}
\end{figure*}

Despite the poor correlation between core mass and sink mass, a Salpeter like mass function is always maintained. We illustrate this graphically in  Figure \ref{rainbowCMF} which shows the cumulative clump mass function of the bound cores. Each point on the graph represents a p-core mass when first bound (our best case scenario) and different mass ranges are denoted with different colours. We now trace the evolution of these cores into sinks, keeping the colours of the sink the same as its parent core. If there was a perfect $1-1$ correlation, then the colour bands would remain distinct. Figure \ref{rainbow} shows the resulting mass functions at subsequent dynamical times. The colours are now well mixed, showing that the p-cores are evolving with variable efficiencies. Nonetheless, the shape of the IMF is maintained throughout, due the the effects of competitive accretion. Therefore, we could say that for a population of cores there is a high probability that a more massive core will form a more massive star, but for a specific object no reliable predictions of final mass can be made. 

In summary, the bound p-cores all form stars, but some are more successful at accreting additional mass than others. This is most likely due to environmental factors, for example, core geometry, surrounding gas reservoir, dynamical interactions or competition from neighbouring proto-stars. This suggests, that in order to understand star-formation, not only the cores of gas must be studied, but also the wider environment in which they form. The correspondence between cores and stars is even more difficult to determine at a snapshot in time when cores are at a variety of evolutionary stages.

\section{Discussion}
\label{sec:discussion}

It has been theorised that the power law clump mass function is formed as a product of supersonic turbulence \citep{Henriksen86,Larson92,Elmegreen97,Klessen01,Padoan02,Hennebelle08}. The core mass function has been observed by many authors (e.g. \citealt{Motte98,Testi98,Johnstone00,Nutter07,Ikeda07}) and bears a remarkable resemblance to the stellar IMF. We now confirm that this profile is also observed in the first bound fragments we have identified in our simulation.

However, the p-cores are not directly comparable to observed cores as they are identified in three dimensions, and use gravitational potential instead of density. Further, it has also been shown that the core properties obtained from applying CLUMPFIND to 2D observations are unreliable \citep{Kainulainen09,Smith08}. Our p-cores are more well defined as they are found from a smooth distribution where all structures are significant. Comparisons between observed core properties and p-core properties must therefore be made cautiously. Nonetheless it is useful to contrast the typical sizes of observed cores to identify which objects are on the most similar size scales to our first bound fragments. Table \ref{cprops} summarises observed properties of density cores in different regions and species from a variety of authors.

\begin{table*}
	\centering
		\caption{A comparison to a small selection of core measurements from the literature. Authors are shown, additional velocities are from * \citet{Lada08}, **\citet{Rosolowsky08} and show the 1D internal velocity dispersion. \citet{Simpson08} is a re-analyses of mm observations in Ophiuchus from various published works. Note the p-cores found in this work are defined differently, as they use 3D potential rather than 2D column density.}
		\begin{tabular}{l c c c c c}
   	         \hline
	         \hline
	          Author & Region & Species & Typical Clump Size & Clump Mass Range & Mean Velocity Dispersion\\
	           & & & [AU] & M$_{\odot}$ & kms$^{1}$ \\
	          \hline
	         \citet{Ikeda07} & OrionA & H13CO+ & $2.9\times10^{4}$ & $2.1-81$&$0.52$  \\ 
		 \citet{Johnstone06} & Orion B & mm cont &$1.4\times10^{4}$ & $0.4 - 30$ & -\\
		 \citet{Alves07} & Ophiuchus & extinction & $\sim 5 \times 10^{4}$ & $0.5 - 28$ & $<0.2$* \\
		 \citet{Simpson08} & Ophiuchus & mm 850mic & $\sim 3 \times 10^{3}$ &$0.1-10$& - \\
		 \citet{Enoch08} & Oph/Per/Serp & 1.1mm cont & $\sim 7\times 10^{3}$ & $0.2-4.8$ & $0.24$** \\
		 \\
		 This work & simulated & potential (bound) & $2.4\times10^{3}$ & $0.2-5$ & $0.16$ \\
		 This work & simulated & potential (composite) &  $3.7\times10^{3}$ & $0.2-10$ & $0.23$ \\
 		\hline
		\end{tabular}
	\label{cprops}
\end{table*}

P-cores are typically smaller than most density cores identified observationally, having average masses of $0.7$ M$_{\odot}$ and radii of $2.4\times10^{3}$ AU when first bound. Their internal 1D velocity dispersions are typically just subsonic. The p-cores are most similar to high resolution observations of cores in nearby molecular clouds (e.g. \citealt{Simpson08,Enoch08}). We expect that the difference between most observed cores and our potential cores is mainly due to resolution. Most surveys, particularly of more distant regions such as Orion, would not be able to resolve cores of the size found here, and as we showed in \citet{Smith08} coarse resolution observations still produce a Salpeter like mass function, but with systematically higher masses.

Although the clump mass function observed from density fluctuations is shown to resemble the stellar initial mass function, there was no requirement for this to be true for the distribution of masses at which cores first become bound. In fact, in the competitive accretion theory of star formation the seeds of gravitational collapse can follow any distribution and the IMF will still be generated through subsequent accretion \citep{Zinnecker82}. That the time integrated bound fragments also follow this distribution further supports the idea that they are sampled from a hierarchical density distribution generated by turbulence. There is some indication that this may not be true at late snapshots in time. Few large diffuse p-cores are formed later in the simulation, which results in a steeper mass function at late epochs.

However, at a snapshot in time the majority ($76\%$) of the p-cores are unbound and may be transitory. This has implications for observational surveys which are necessarily at a snapshot in time, as the vast majority of objects have either not reached the point of collapse or will not form stars at all. The case for a 1-1 correlation between core mass and stellar mass can not hold in these cases. There is also the further complication that our p-cores contain substructure that could be further identified as individual objects if one were performing a density decomposition (depending on the resolution of the data). Naturally, these yet smaller objects are even less bound than their parent p-core. The high value of unbound pre-stellar p-cores in our analysis is surprising because as we identify them using gravitational potential this force must be significant. There is observational evidence for unbound cores. \citet{Lada08} find that the majority of cores found from dust extinction in the Pipe nebula are unbound, and \citet{Andre07} find $25\%$ of their cores in Ophiuchus to be unbound.

The p-cores showed significant density sub-structure, despite being identified from a smooth potential well. If a traditional clump finding algorithm using emission or density had been used to identify them, they would have been split into smaller objects. The p-cores were also only quasi-spherical, and their central peak in potential did not necessarily correspond with their center of mass, as they were not relaxed. This suggests that using smooth symmetrical models to study core collapse (such as Bonnor-Ebert spheres) can at best only be a first approximation.

Unfortunately, our simulation does not include magnetic fields, which can act against collapse on some length scales \citep{Hennebelle06}. \citet{Price08} found that magnetic fields introduce voids into molecular cloud structure and reduce subsequent accretion. We also do not have radiative transfer in our simulation. However, as we are considering the cores early in their evolution before a protostar is formed, and accretion luminosity only affects scales of a few $100$ AU \citep{Krumholz07b, Bate09}, this should not be a significant issue.

Due to the similarities between the Stellar IMF and the clump mass function it has been proposed that the mass of stellar systems is directly related to the clump they formed from (e.g. \citealt{Motte98,Alves07}). However, when we traced the total mass in sink particles formed from our p-cores at a snapshot in time, as would be visible observationally, the correspondence was effectively non-existent. A stronger correlation between core masses and total sink mass was found when the p-core masses were recorded at the time when they were first bound. Further accretion increases the dispersion in this relation and introduces an offset. 

Throughout the simulation the most massive bound core had a mass of only $6.35$ M$_{\odot}$, whereas the most massive sink at the end of the simulation had a mass of $27.97$ M$_{\odot}$. We did not find any p-core which could form a massive star simply from its own material when first bound. All the bound cores were at best only marginally supersonic and hence were not supported by turbulence, contrary to the massive turbulent cores predicted by \citet{McKee03}. Note, however, that the lack of feedback and magnetic fields in these simulations may limit the ability to sustain turbulence in cores.

In most cases gas from outside the region initially bound was accreted by the sinks formed from the p-core after only $6\times10^{4}$ yrs. This subsequent accretion means that the environment surrounding the p-core is also important for its future evolution, for example whether it is surrounded by a large gas reservoir or is in a narrow filament.

There is a higher probability that a more massive p-core will form a larger total sink mass, but for an individual p-core no accurate prediction can be made. The link between core mass and sink mass is poor because there are a number of additional environmental factors beyond the bound mass which will affect the evolution of the core, for example the shape of the core and the distribution of gas it is embedded within. Despite the poor correspondence between p-core masses and sinks, the shape of the mass function obtained from the p-cores as they evolve into sinks always resembles the stellar IMF.

\section{Conclusions}

We identified the earliest fragmentation in molecular clouds using the distribution of gravitational potential rather than density. This is a smoother distribution, due to gravity being a long range force, and it allows us to assign boundaries with clearer physical meanings. We called our fragments `p-cores', and traced their evolution with time. If they became bound, we identified their mass when ($E_{p}>E_{k}+E_{therm}$) for the first time, and we related this to the mass in sink particles formed from them. This allowed us to identify the scale of the initial fragmentation in molecular clouds, and trace its evolution towards the stellar IMF. Our main conclusions are as follows:

\begin{enumerate}
\item The time-integrated mass function of just-bound gravitational potential `p-cores' resembles the stellar initial mass function in a similar manner to that of gas density cores
\item The bound p-cores are most similar to the very smallest density cores currently observable ( an average of $0.7$ M$_{\odot}$ with a radius of $2300$ AU).
\item P-cores exhibit significant density substructure.
\item There is a poor correlation between p-core mass and the total mass in sinks formed from them. This is particularly true when the p-cores are recorded at a snap-shot in time, but still holds when the mass is recorded when they are first bound.
\item The sink particles formed from the p-cores accrete from beyond the region initially bound. This means that the surrounding environment of the core also has an effect on the star formation in our simulation.
\item Despite the poor correlation between p-core mass and sink masses the sink IMF always resembles the stellar IMF.
\end{enumerate}

\section{Acknowledgements}
We would like to thank Jane Greaves, Jaime Pineda, Alyssa Goodman, Ralf Klessen, Derek Ward-Thompson, David Nutter and Robert Smith for useful discussions. We would also like to thank the referee, Richard Durisen, for his constructive comments. P.C.C. acknowledges support by the Deutsche 
Forschungsgemeinschaft (DFG) under grant KL 1358/5 and via the 
Sonderforschungsbereich (SFB) SFB 439, Galaxien im fr\"uhen Universum. Lastly, we would like to acknowledge the Scottish Universities Physics Alliance (SUPA) for providing the computing facilities required for this work. 

\bibliography{Bibliography}

\begin{thebibliography}{}

\bibitem[\protect\citeauthoryear{{Alves}, {Lombardi} \& {Lada}}{{Alves}
  et~al.}{2007}]{Alves07}
{Alves} J.,  {Lombardi} M.,    {Lada} C.~J.,  2007, \aap, 462, L17

\bibitem[\protect\citeauthoryear{{Andr{\'e}}, {Belloche}, {Motte} \&
  {Peretto}}{{Andr{\'e}} et~al.}{2007}]{Andre07}
{Andr{\'e}} P.,  {Belloche} A.,  {Motte} F.,    {Peretto} N.,  2007, \aap, 472,
  519

\bibitem[\protect\citeauthoryear{{Ballesteros-Paredes}, {Klessen} \&
  {V{\'a}zquez-Semadeni}}{{Ballesteros-Paredes}
  et~al.}{2003}]{Ballesteros-Paredes03}
{Ballesteros-Paredes} J.,  {Klessen} R.~S.,    {V{\'a}zquez-Semadeni} E.,
  2003, \apj, 592, 188

\bibitem[\protect\citeauthoryear{{Bate}}{{Bate}}{2009}]{Bate09}
{Bate} M.~R.,  2009, \mnras, 392, 1363

\bibitem[\protect\citeauthoryear{{Bate}, {Bonnell} \& {Price}}{{Bate}
  et~al.}{1995}]{Bate95}
{Bate} M.~R.,  {Bonnell} I.~A.,    {Price} N.~M.,  1995, \mnras, 277, 362

\bibitem[\protect\citeauthoryear{{Bate} \& {Burkert}}{{Bate} \&
  {Burkert}}{1997}]{Bate97}
{Bate} M.~R.,  {Burkert} A.,  1997, \mnras, 288, 1060

\bibitem[\protect\citeauthoryear{{Benz}}{{Benz}}{1990}]{Benz90}
{Benz} W.,  1990, in {Buchler} J.~R.,  ed., Numerical Modelling of Nonlinear
  Stellar Pulsations Problems and Prospects {Smooth Particle Hydrodynamics - a
  Review}.
Kluwer Academic Publishers, Dordrecht, The Netherlands, pp 269--+

\bibitem[\protect\citeauthoryear{{Bergin} \& {Tafalla}}{{Bergin} \&
  {Tafalla}}{2007}]{Bergin07}
{Bergin} E.~A.,  {Tafalla} M.,  2007, \araa, 45, 339

\bibitem[\protect\citeauthoryear{{Bonnell} \& {Bate}}{{Bonnell} \&
  {Bate}}{2006}]{Bonnell06}
{Bonnell} I.~A.,  {Bate} M.~R.,  2006, \mnras, 370, 488

\bibitem[\protect\citeauthoryear{{Bonnell}, {Vine} \& {Bate}}{{Bonnell}
  et~al.}{2004}]{Bonnell04}
{Bonnell} I.~A.,  {Vine} S.~G.,    {Bate} M.~R.,  2004, \mnras, 349, 735

\bibitem[\protect\citeauthoryear{{Bonnor}}{{Bonnor}}{1956}]{Bonnor56}
{Bonnor} W.~B.,  1956, \mnras, 116, 351

\bibitem[\protect\citeauthoryear{{Chabrier}}{{Chabrier}}{2003}]{Chabrier03}
{Chabrier} G.,  2003, \pasp, 115, 763

\bibitem[\protect\citeauthoryear{{Clark} \& {Bonnell}}{{Clark} \&
  {Bonnell}}{2005}]{Clark05}
{Clark} P.~C.,  {Bonnell} I.~A.,  2005, \mnras, 361, 2

\bibitem[\protect\citeauthoryear{{Clark}, {Klessen} \& {Bonnell}}{{Clark}
  et~al.}{2007}]{Clark07}
{Clark} P.~C.,  {Klessen} R.~S.,    {Bonnell} I.~A.,  2007, \mnras, 379, 57

\bibitem[\protect\citeauthoryear{{Dib}, {Kim}, {V{\'a}zquez-Semadeni},
  {Burkert} \& {Shadmehri}}{{Dib} et~al.}{2007}]{Dib07}
{Dib} S.,  {Kim} J.,  {V{\'a}zquez-Semadeni} E.,  {Burkert} A.,    {Shadmehri}
  M.,  2007, \apj, 661, 262

\bibitem[\protect\citeauthoryear{{Dobbs}, {Bonnell} \& {Clark}}{{Dobbs}
  et~al.}{2005}]{Dobbs05}
{Dobbs} C.~L.,  {Bonnell} I.~A.,    {Clark} P.~C.,  2005, \mnras, 360, 2

\bibitem[\protect\citeauthoryear{{Dubinski}, {Narayan} \&
  {Phillips}}{{Dubinski} et~al.}{1995}]{Dubinski95}
{Dubinski} J.,  {Narayan} R.,    {Phillips} T.~G.,  1995, \apj, 448, 226

\bibitem[\protect\citeauthoryear{{Ebert}}{{Ebert}}{1955}]{Ebert55}
{Ebert} R.,  1955, Zeitschrift fur Astrophysik, 37, 217

\bibitem[\protect\citeauthoryear{{Elmegreen}}{{Elmegreen}}{1997}]{Elmegreen97}
{Elmegreen} B.~G.,  1997, \apj, 486, 944

\bibitem[\protect\citeauthoryear{{Elmegreen}, {Klessen} \&
  {Wilson}}{{Elmegreen} et~al.}{2008}]{Elmegreen08}
{Elmegreen} B.~G.,  {Klessen} R.~S.,    {Wilson} C.~D.,  2008, \apj, 681, 365

\bibitem[\protect\citeauthoryear{{Elmegreen} \& {Mathieu}}{{Elmegreen} \&
  {Mathieu}}{1983}]{Elmegreen83}
{Elmegreen} B.~G.,  {Mathieu} R.~D.,  1983, \mnras, 203, 305

\bibitem[\protect\citeauthoryear{{Enoch}, {Evans} II, {Sargent}, {Glenn},
  {Rosolowsky} \& {Myers}}{{Enoch} et~al.}{2008}]{Enoch08}
{Enoch} M.~L.,  {Evans} II N.~J.,  {Sargent} A.~I.,  {Glenn} J.,  {Rosolowsky}
  E.,    {Myers} P.,  2008, \apj, 684, 1240

\bibitem[\protect\citeauthoryear{{Gingold} \& {Monaghan}}{{Gingold} \&
  {Monaghan}}{1983}]{Gingold83}
{Gingold} R.~A.,  {Monaghan} J.~J.,  1983, \mnras, 204, 715

\bibitem[\protect\citeauthoryear{{Goodman}, {Barranco}, {Wilner} \&
  {Heyer}}{{Goodman} et~al.}{1998}]{Goodman98}
{Goodman} A.~A.,  {Barranco} J.~A.,  {Wilner} D.~J.,    {Heyer} M.~H.,  1998,
  \apj, 504, 223

\bibitem[\protect\citeauthoryear{{Goodwin}, {Nutter}, {Kroupa}, {Ward-Thompson}
  \& {Whitworth}}{{Goodwin} et~al.}{2008}]{Goodwin08}
{Goodwin} S.~P.,  {Nutter} D.,  {Kroupa} P.,  {Ward-Thompson} D.,
  {Whitworth} A.~P.,  2008, \aap, 477, 823

\bibitem[\protect\citeauthoryear{{Heitsch}, {Mac Low} \& {Klessen}}{{Heitsch}
  et~al.}{2001}]{Heitsch01}
{Heitsch} F.,  {Mac Low} M.-M.,    {Klessen} R.~S.,  2001, \apj, 547, 280

\bibitem[\protect\citeauthoryear{{Hennebelle} \& {Chabrier}}{{Hennebelle} \&
  {Chabrier}}{2008}]{Hennebelle08}
{Hennebelle} P.,  {Chabrier} G.,  2008, ArXiv e-prints, 805

\bibitem[\protect\citeauthoryear{{Hennebelle} \& {Passot}}{{Hennebelle} \&
  {Passot}}{2006}]{Hennebelle06}
{Hennebelle} P.,  {Passot} T.,  2006, \aap, 448, 1083

\bibitem[\protect\citeauthoryear{{Henriksen}}{{Henriksen}}{1986}]{Henriksen86}
{Henriksen} R.~N.,  1986, \apj, 310, 189

\bibitem[\protect\citeauthoryear{{Ikeda}, {Sunada} \& {Kitamura}}{{Ikeda}
  et~al.}{2007}]{Ikeda07}
{Ikeda} N.,  {Sunada} K.,    {Kitamura} Y.,  2007, \apj, 665, 1194

\bibitem[\protect\citeauthoryear{{Jappsen}, {Klessen}, {Larson}, {Li} \& {Mac
  Low}}{{Jappsen} et~al.}{2005}]{Jappsen05}
{Jappsen} A.-K.,  {Klessen} R.~S.,  {Larson} R.~B.,  {Li} Y.,    {Mac Low}
  M.-M.,  2005, \aap, 435, 611

\bibitem[\protect\citeauthoryear{Jeans}{Jeans}{1902}]{Jeans1902}
Jeans J.~H.,  1902, Philosophical Transactions of the Royal Society of London.
  Series A, Containing Papers of a Mathematical or Physical Character, 199, 1

\bibitem[\protect\citeauthoryear{{Johnstone}, {Matthews} \&
  {Mitchell}}{{Johnstone} et~al.}{2006}]{Johnstone06}
{Johnstone} D.,  {Matthews} H.,    {Mitchell} G.~F.,  2006, \apj, 639, 259

\bibitem[\protect\citeauthoryear{{Johnstone}, {Wilson}, {Moriarty-Schieven},
  {Joncas}, {Smith}, {Gregersen} \& {Fich}}{{Johnstone}
  et~al.}{2000}]{Johnstone00}
{Johnstone} D.,  {Wilson} C.~D.,  {Moriarty-Schieven} G.,  {Joncas} G.,
  {Smith} G.,  {Gregersen} E.,    {Fich} M.,  2000, \apj, 545, 327

\bibitem[\protect\citeauthoryear{{Kainulainen}, {Lada}, {Rathborne} \&
  {Alves}}{{Kainulainen} et~al.}{2009}]{Kainulainen09}
{Kainulainen} J.,  {Lada} C.~J.,  {Rathborne} J.~M.,    {Alves} J.~F.,  2009,
  ArXiv e-prints

\bibitem[\protect\citeauthoryear{{Klessen}}{{Klessen}}{2001}]{Klessen01}
{Klessen} R.~S.,  2001, \apj, 556, 837

\bibitem[\protect\citeauthoryear{{Klessen} \& {Burkert}}{{Klessen} \&
  {Burkert}}{2000}]{Klessen00}
{Klessen} R.~S.,  {Burkert} A.,  2000, \apjs, 128, 287

\bibitem[\protect\citeauthoryear{{Krumholz}, {Klein} \& {McKee}}{{Krumholz}
  et~al.}{2007}]{Krumholz07b}
{Krumholz} M.~R.,  {Klein} R.~I.,    {McKee} C.~F.,  2007, \apj, 656, 959

\bibitem[\protect\citeauthoryear{{Lada}, {Muench}, {Rathborne}, {Alves} \&
  {Lombardi}}{{Lada} et~al.}{2008}]{Lada08}
{Lada} C.~J.,  {Muench} A.~A.,  {Rathborne} J.,  {Alves} J.~F.,    {Lombardi}
  M.,  2008, \apj, 672, 410

\bibitem[\protect\citeauthoryear{{Lada}}{{Lada}}{1992}]{Lada_E92}
{Lada} E.~A.,  1992, \apjl, 393, L25

\bibitem[\protect\citeauthoryear{{Larson}}{{Larson}}{1973}]{Larson73}
{Larson} R.~B.,  1973, \mnras, 161, 133

\bibitem[\protect\citeauthoryear{{Larson}}{{Larson}}{1992}]{Larson92}
{Larson} R.~B.,  1992, \mnras, 256, 641

\bibitem[\protect\citeauthoryear{{Larson}}{{Larson}}{2005}]{Larson05}
{Larson} R.~B.,  2005, \mnras, 359, 211

\bibitem[\protect\citeauthoryear{{Masunaga} \& {Inutsuka}}{{Masunaga} \&
  {Inutsuka}}{2000}]{Masunaga00}
{Masunaga} H.,  {Inutsuka} S.-i.,  2000, \apj, 531, 350

\bibitem[\protect\citeauthoryear{{Matzner} \& {McKee}}{{Matzner} \&
  {McKee}}{2000}]{Matzner00}
{Matzner} C.~D.,  {McKee} C.~F.,  2000, \apj, 545, 364

\bibitem[\protect\citeauthoryear{{McKee} \& {Tan}}{{McKee} \&
  {Tan}}{2003}]{McKee03}
{McKee} C.~F.,  {Tan} J.~C.,  2003, \apj, 585, 850

\bibitem[\protect\citeauthoryear{{Monaghan}}{{Monaghan}}{1992}]{Monaghan92}
{Monaghan} J.~J.,  1992, \araa, 30, 543

\bibitem[\protect\citeauthoryear{{Motte}, {Andre} \& {Neri}}{{Motte}
  et~al.}{1998}]{Motte98}
{Motte} F.,  {Andre} P.,    {Neri} R.,  1998, \aap, 336, 150

\bibitem[\protect\citeauthoryear{{Myers}}{{Myers}}{2008}]{Myers08}
{Myers} P.~C.,  2008, ArXiv e-prints

\bibitem[\protect\citeauthoryear{{Myers} \& {Gammie}}{{Myers} \&
  {Gammie}}{1999}]{Myers99}
{Myers} P.~C.,  {Gammie} C.~F.,  1999, \apjl, 522, L141

\bibitem[\protect\citeauthoryear{{Nutter} \& {Ward-Thompson}}{{Nutter} \&
  {Ward-Thompson}}{2007}]{Nutter07}
{Nutter} D.,  {Ward-Thompson} D.,  2007, \mnras, 374, 1413

\bibitem[\protect\citeauthoryear{{Padoan}, {Cambr{\'e}sy}, {Juvela}, {Kritsuk},
  {Langer} \& {Norman}}{{Padoan} et~al.}{2006}]{Padoan06}
{Padoan} P.,  {Cambr{\'e}sy} L.,  {Juvela} M.,  {Kritsuk} A.,  {Langer} W.~D.,
    {Norman} M.~L.,  2006, \apj, 649, 807

\bibitem[\protect\citeauthoryear{{Padoan} \& {Nordlund}}{{Padoan} \&
  {Nordlund}}{2002}]{Padoan02}
{Padoan} P.,  {Nordlund} {\AA}.,  2002, \apj, 576, 870

\bibitem[\protect\citeauthoryear{{Pineda}, {Caselli} \& {Goodman}}{{Pineda}
  et~al.}{2008}]{Pineda08}
{Pineda} J.~E.,  {Caselli} P.,    {Goodman} A.~A.,  2008, \apj, 679, 481

\bibitem[\protect\citeauthoryear{{Price} \& {Bate}}{{Price} \&
  {Bate}}{2008}]{Price08}
{Price} D.~J.,  {Bate} M.~R.,  2008, \mnras, 385, 1820

\bibitem[\protect\citeauthoryear{{Rosolowsky}, {Pineda}, {Kauffmann} \&
  {Goodman}}{{Rosolowsky} et~al.}{2008}]{Rosolowsky08}
{Rosolowsky} E.~W.,  {Pineda} J.~E.,  {Kauffmann} J.,    {Goodman} A.~A.,
  2008, \apj, 679, 1338

\bibitem[\protect\citeauthoryear{{Salpeter}}{{Salpeter}}{1955}]{Salpeter55}
{Salpeter} E.~E.,  1955, \apj, 121, 161

\bibitem[\protect\citeauthoryear{{Schnee}, {Li}, {Goodman} \&
  {Sargent}}{{Schnee} et~al.}{2008}]{Schnee08}
{Schnee} S.,  {Li} J.,  {Goodman} A.~A.,    {Sargent} A.~I.,  2008, \apj, 684,
  1228

\bibitem[\protect\citeauthoryear{{Shu}}{{Shu}}{1977}]{Shu77}
{Shu} F.~H.,  1977, \apj, 214, 488

\bibitem[\protect\citeauthoryear{{Shu}, {Lizano}, {Ruden} \& {Najita}}{{Shu}
  et~al.}{1988}]{Shu88}
{Shu} F.~H.,  {Lizano} S.,  {Ruden} S.~P.,    {Najita} J.,  1988, \apjl, 328,
  L19

\bibitem[\protect\citeauthoryear{{Silk}}{{Silk}}{1995}]{Silk95}
{Silk} J.,  1995, \apjl, 438, L41

\bibitem[\protect\citeauthoryear{{Simpson}, {Nutter} \&
  {Ward-Thompson}}{{Simpson} et~al.}{2008}]{Simpson08}
{Simpson} R.~J.,  {Nutter} D.,    {Ward-Thompson} D.,  2008, ArXiv e-prints

\bibitem[\protect\citeauthoryear{{Smith}, {Clark} \& {Bonnell}}{{Smith}
  et~al.}{2008}]{Smith08}
{Smith} R.~J.,  {Clark} P.~C.,    {Bonnell} I.~A.,  2008, \mnras, 391, 1091

\bibitem[\protect\citeauthoryear{{Swift} \& {Williams}}{{Swift} \&
  {Williams}}{2008}]{Swift08}
{Swift} J.~J.,  {Williams} J.~P.,  2008, \apj, 679, 552

\bibitem[\protect\citeauthoryear{{Tachihara}, {Onishi}, {Mizuno} \&
  {Fukui}}{{Tachihara} et~al.}{2002}]{Tachihara02}
{Tachihara} K.,  {Onishi} T.,  {Mizuno} A.,    {Fukui} Y.,  2002, \aap, 385,
  909

\bibitem[\protect\citeauthoryear{{Testi} \& {Sargent}}{{Testi} \&
  {Sargent}}{1998}]{Testi98}
{Testi} L.,  {Sargent} A.~I.,  1998, \apjl, 508, L91

\bibitem[\protect\citeauthoryear{{Tilley} \& {Pudritz}}{{Tilley} \&
  {Pudritz}}{2007}]{Tilley07}
{Tilley} D.~A.,  {Pudritz} R.~E.,  2007, \mnras, 382, 73

\bibitem[\protect\citeauthoryear{{Ward-Thompson}, {Scott}, {Hills} \&
  {Andre}}{{Ward-Thompson} et~al.}{1994}]{Ward-Thompson94}
{Ward-Thompson} D.,  {Scott} P.~F.,  {Hills} R.~E.,    {Andre} P.,  1994,
  \mnras, 268, 276

\bibitem[\protect\citeauthoryear{{Williams}, {de Geus} \& {Blitz}}{{Williams}
  et~al.}{1994}]{Williams94}
{Williams} J.~P.,  {de Geus} E.~J.,    {Blitz} L.,  1994, \apj, 428, 693

\bibitem[\protect\citeauthoryear{{Young}, {Shirley}, {Evans} II \&
  {Rawlings}}{{Young} et~al.}{2003}]{Young03}
{Young} C.~H.,  {Shirley} Y.~L.,  {Evans} II N.~J.,    {Rawlings} J.~M.~C.,
  2003, \apjs, 145, 111

\bibitem[\protect\citeauthoryear{{Zinnecker}}{{Zinnecker}}{1982}]{Zinnecker82}
{Zinnecker} H.,  1982, New York Academy Sciences Annals, 395, 226

\end{thebibliography}

\label{lastpage}

\end{document}